\documentclass[longbibliography,twocolumn,superscriptaddress, aps]{revtex4-1}
\usepackage{amsmath,amssymb,graphicx,color,bm}
\usepackage[hidelinks]{hyperref}

\begin{document}

\title{Intersubband plasmon excitations in doped carbon
  nanotubes}

\author{Daria~Satco}
\email{daria.satco@skoltech.ru}
\affiliation{Skolkovo Institute of Science and Technology,
  Moscow 143026, Russia}
\affiliation{Department of Physics, Tohoku
  University, Sendai 980-8578, Japan}

\author{Ahmad~R.~T.~Nugraha}
\email{nugraha@flex.phys.tohoku.ac.jp}
\affiliation{Department of Physics, Tohoku University, Sendai
  980-8578, Japan}

\author{M.~Shoufie~Ukhtary}
\affiliation{Department of Physics, Tohoku University, Sendai
  980-8578, Japan}

\author{Daria~Kopylova}
\affiliation{Skolkovo Institute of Science and Technology,
  Moscow 143026, Russia}

\author{Albert~G.~Nasibulin}
\affiliation{Skolkovo Institute of Science and Technology,
  Moscow 143026, Russia}
\affiliation{Department of Applied Physics, Aalto University,
  Aalto 00076, Finland}

\author{Riichiro~Saito}
\affiliation{Department of Physics, Tohoku University, Sendai
  980-8578, Japan}


\begin{abstract}
We theoretically investigate intersubband plasmon excitations in doped
single wall carbon nanotubes (SWNTs) by examining the dependence of
plasmon frequency on the nanotube diameter, chirality, and Fermi
energy.  The intersubband plasmons can be excited by light with
polarization perpendicular to the nanotube axis and thus the plasmon
excitations corresponds to optical transitions between the two
different subbands, which are sensitive to the Fermi energy.  In every
SWNT, this mechanism leads to the emergence of the optical absorption
peak at the plasmon frequency for a given Fermi energy, $E_F$.  The
plasmon frequencies calculated for many SWNTs with diameter $d_t <
2$~nm exhibit a dependence on $(1/d_t)^{0.7}$ and the frequencies are
further affected by Fermi energy as $E_F^{0.25}$.  With this
knowledge, it is possible to develop a map of intersubband plasmon
excitations in doped SWNTs that could be useful to quickly estimate
the doping level and also be an alternative way to characterize
nanotube chirality.
\end{abstract}

\maketitle

\section{Introduction}

For many years, single wall carbon nanotubes (SWNTs) have been an
important platform to study optical properties of one-dimensional (1D)
materials, especially due to their geometry-dependent optical
absorption~\cite{Kataura1999,Saito2000,Weisman2002,Weisman2003} and
also due to their potential applications for optoelectronic
devices~\cite{Avouris2006,Avouris2008,Kaskela2010,Tsapenko2018}.  Of
the wide interests in the optical properties of SWNTs, a particular
problem of the doping effects on the absorption of linearly polarized
light is worth investigating.  So far, previous studies have confirmed
that \emph{undoped} SWNTs absorb only light with polarization parallel
to the nanotube
axis~\cite{Ajiki94,Hwang2000,Jiang2004,Murakami2005,Li2001}, so that
when the light polarization is perpendicular to the nanotube axis the
undoped SWNTs do not show any absorption peak due to the
depolarization effect~\cite{Uryu2006,Jiang2004}.  The optical
absorption in the case of parallel polarization can be understood in
terms of the $E_{ii}$ \emph{interband} excitations from the \emph{i}th
valence to the \emph{i}th conduction energy subbands, either in
single-particle~\cite{Ajiki94,Saito2000,Jiang2004} or excitonic
picture~\cite{Spataru2004,Wang2005,Dukovic2005,Jiang2007}.  On the
other hand, much uncertainty still exists about what happens in the
case of \emph{doped} SWNTs for the linearly polarized light.

Recently, Sasaki \emph{et al.} suggested that doped (undoped) SWNT
absorb light with polarization perpendicular (parallel) to the
nanotube axis~\cite{sasaki16-interplasmon,sasaki18-plasmonpol}.
Furthermore, Yanagi~\emph{et al.}~\cite{yanagi18-isbp} experimentally
gave evidence that the doped SWNTs absorb light with the perpendicular
polarization within the near-infrared range of a photon energy
($\sim$$0.8$--$1.2$~eV).  This energy range is similar to that when
undoped SWNTs absorb light with the parallel polarization.  Senga
\emph{et al.}  showed consistent absorption peaks for isolated
metallic SWNTs that are unintendedly doped on the TEM supporting grid
during electron energy-loss spectroscopy (EELS)
measurement~\cite{Senga2016,senga18-eels}.  Yanagi~\emph{et al.}
proposed that the absorption peaks are related with
\emph{intersubband} plasmon excitations~\cite{yanagi18-isbp}, i.e.,
the optical transitions with energies
$E_{ij}$ occur collectively between two electronic subbands
$i$ and
$j$ as a response to the perpendicularly polarized light.  Unlike the
\emph{interband} excitations
$E_{ii}$ which take place from the valence to the conduction bands,
the \emph{intersubband} plasmon excitations
$E_{ij}$ occur within the conduction band or within the valence band.

It should be noted that in the EELS experiment by Senga~\emph{et al.}
we can also see another plasmonic peak around $6~\mathrm{eV}$, the
so-called $\pi$ plasmon, which is \emph{not} excited by light with
perpendicular polarization but with parallel
polarization~\cite{Senga2016,senga18-eels}.  Observations of the
$\pi$-plasmons in SWNTs~\cite{Kuzuo1992,Kuzuo1994} or any graphitic
materials~\cite{Papagno1983,Liou2014,Hu2014}, either doped or undoped,
are quite common in the earlier EELS experiments and the peaks are
assigned unambiguously.  Lin and Shung in two decades ago
theoretically explained the origin of $\pi$ plasmons in the SWNTs as a
result of collective interband excitations of the $\pi$-band
electrons~\cite{Lin1994,Lin1996}.  On the other hand, the theory for
plasmons excited in doped SWNTs with perpendicularly polarized light
is just available recently by Sasaki \emph{et
  al.}~\cite{sasaki16-interplasmon,sasaki18-plasmonpol} and Garcia de
Abajo~\cite{Abajo2014}, in which they discussed how the plasmon
frequency $(\omega_p$) in a doped SWNT depends on its diameter ($d_t$)
and Fermi energy ($E_F$).  However, the dependence of $\omega_p$
on $d_t$ and $E_F$ was analyzed within the Drude model, which is not
relevant to \emph{intersubband} transitions but it deals with
\emph{intrasubband} transitions.  In this sense, there is a necessity
to properly describe the intersubband plasmons in the doped SWNTs for
any SWNT structure or chirality.

In this work, we show our calculation of plasmon frequencies for the
doped SWNTs as a function of diameter and the Fermi energy,
considering all SWNTs with different chiralities in the range of
$0.5 < d_t < 2$~nm.  The calculated plasmon frequencies exhibit a
diameter dependence of $(1/d_t)^{0.7}$ and are further dependent on
the Fermi energy as $E_F^{0.25}$.  This scaling of plasmon frequency
differs with that predicted by the Drude model,
$\omega_p \propto
(E_F/d_t)^{0.5}$~\cite{sasaki16-interplasmon,Abajo2014}, hence
indicating the difference of the \emph{intersubband} transitions
(current work) from the \emph{intrasubband} transitions (the Drude
model).  We further consider optical absorption at the plasmon
frequencies caused by intersubband transitions within the conduction
and valence bands, corresponding to $E_F > 0$ and $E_F < 0$,
respectively.  We find that the most dominant plasmonic transition,
which we label as $\mathrm{P}_{ij}$ at a certain energy $E_{ij}$
(following the notation introduced by Bondarev~\cite{Bondarev2012} for
the \emph{interband} plasmon at $E_{ii}$), changes with Fermi energy
from a $\mathrm{P}_{ij}$ to another $\mathrm{P}_{i'j'}$.  For the
smaller (larger) nanotube diameter, we need higher (lower) $E_F$ to
excite the plasmon.  Using the fitting formula for the plasmon
frequency provided in this paper, one can estimate the Fermi energy in
the doped SWNTs by means of optical spectroscopy, as well as EELS.
Furthermore, experimentalists can also search for intersubband
plasmons in isolated SWNTs with various chiralities, not only limited
to SWNTs bundles.

The rest of this paper is organized as follows.  In Sec.~\ref{sec:th},
we describe how to calculate the plasmon frequency for a given SWNT
starting from the dielectric function of the SWNT.  The complex
dielectric function in this work is calculated within the
self-consistent-field approach by considering dipole approximation for
optical matrix elements, from which there exist selection rules for
different light polarization.  In Sec.~\ref{sec:res}, we discuss the
main results of intersubband plasmon frequencies, including the
opportunity to map them into a unified picture of
$\omega_p \propto (E_F^{0.25}/d_t^{0.7})$.  We justify the fitting by
means of graphene plasmon dispersion, considering the model of the
rolled graphene sheet for a SWNT.  Finally, we give conclusions and
future perspectives in Sec.~\ref{sec:con}.

\section{Theoretical methods} \label{sec:th}
\subsection{Defining plasmons from dielectric function}
\label{subsec:diel}

We consider a SWNT subjected to perturbation by light whose vector
potential, electric field, and magnetic field are denoted by
$\mathbf{A}$, $\mathbf{E}$, and $\mathbf{B}$, respectively.  The
vector potential of the electric field of incident light at the
position of $\mathbf{r}$ and time $t$ is given by:
\begin{equation}
\label{vect_pot}
\mathbf{A}(\mathbf{r}, t)
= A_0 \mathbf{n} \cos (\mathbf{q} \cdot \mathbf{r} - \omega t),
\end{equation}
where $A_0$, $\omega$, $\mathbf{q}$, and $\mathbf{n}$ denote the
vector potential amplitude, angular frequency, wave vector in the
direction of propagation, and unit vector of polarization direction,
respectively.  The magnetic and electric fields are related with
$\mathbf{A}$ by $\mathbf{E}(\mathbf{r},t) = -d\mathbf{A}/dt$ and
$\mathbf{B}(\mathbf{r},t) = \mathbf{\nabla} \times \mathbf{A}$,
respectively.  These quantities are important in the calculation of
optical matrix elements, as derived in details in
Appendix~\ref{app:selection}.

\begin{figure}[t!]
  \centering \includegraphics[clip,width=8cm]{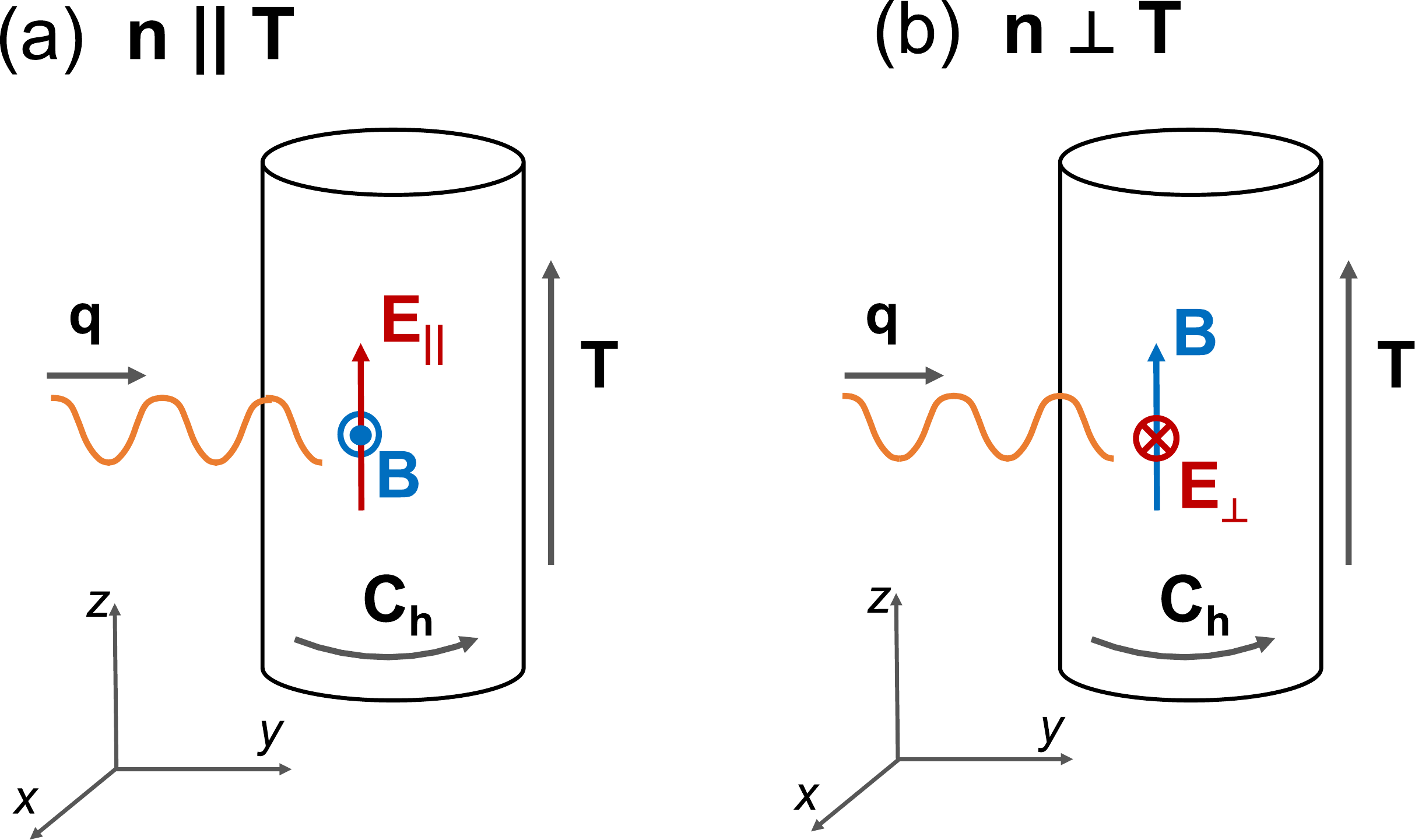}
  \caption{\label{fig:1} Two geometries of propagating linearly
    polarized light with electrical field being polarized in (a)
    parallel and (b) perpendicular directions with respect to the
    nanotube axis.}
\end{figure}

We will discuss two cases of $\mathbf{n}$: parallel and perpendicular
to the nanotube axis, shown in Fig.~\ref{fig:1}.  We refer to the two
cases as the parallel polarization and perpendicular polarization.
The nanotube axis is denoted by the translational vector $\mathbf{T}$
in three dimension as shown in Fig.~\ref{fig:1}(a) for $\mathbf{n}
\parallel \mathbf{T}$ and Fig.~\ref{fig:1}(b) for $\mathbf{n} \perp
\mathbf{T}$.  If we imagine the SWNT as a rolled-up graphene sheet,
the nanotube axis in the unrolled sheet is always perpendicular to the
chiral vector $\mathbf{C}_h$, thus the unit cell of the SWNT is
defined by the rectangular whose boundaries are $\mathbf{C}_h$ and
$\mathbf{T}$~\cite{saito1998physical}.  The chiral vector
$\mathbf{C}_h$ in the basis of two-dimensional (2D) lattice vectors of
graphene uniquely identifies the SWNT structure by $\mathbf{C}_h =
(n,m)$, where the set of integers $(n,m)$ is known as the chirality.

In both optical spectroscopy and EELS, plasmons are observed as
prominent peaks in the spectra.  The intensity of optical absorption
is proportional to
$\mathrm{Re} (\sigma/ \varepsilon)~$\cite{weber1999intersubband},
where $\sigma$ and $\varepsilon$ are, respectively, optical
conductivity and dielectric function as a function of light frequency
$\omega$.  Note that the dielectric function in the optical absorption
accounts for the depolarization effect, which means that the screening
of the external electrical field is included in the calculation of
optical absorption for both perpendicular and parallel polarizations
of light.  Indeed, the depolarization effect is essential for
explaining the anisotropy of optical absorption in
SWNTs~\cite{Ajiki1993,Uryu2006,Nakanishi2009}.  On the other hand, the
intensity of EELS is proportional to the energy loss-function,
$\mathrm{Im}(-1/\varepsilon)$~\cite{Ritchie1957,Raether1980}, that
describes the excitation spectrum of solid by inelastic scattering of
electrons at small angles.  The plasmon peaks originate from zero
points of the real part of $\varepsilon(\omega)$, i.e.,
$\mathrm{Re}[\varepsilon(\omega)] = 0$, followed by a relatively small
value of its imaginary part, $\mathrm{Im}[\varepsilon(\omega)]$, in
comparison with the maximum of $\mathrm{Im}[\varepsilon(\omega)]$.

According to the Maxwell equations, the optical conductivity $\sigma$
is related to the dielectric function $\varepsilon$ as follows:
\begin{equation}
\label{conductivity}
\varepsilon(\omega) = 1 + i \frac{4 \pi \sigma(\omega)}{\omega L \varepsilon_s},
\end{equation}
where $\varepsilon_s$ is surrounding dielectric permittivity
($\varepsilon_s = 2$ for SWNT film~\cite{Igarashi2015}) and $L$ is
effective thickness of the material ($L = d_t$ for SWNT).  We
calculate $\varepsilon(\omega)$ within the self-consistent-field
approach in the following form \cite{Ehrenreich1959, Lin1994}:
\begin{align}
\label{epsilon}
\varepsilon(\omega) = \varepsilon_0 + &\bigg[\frac{8 \pi e^2}{\hbar \omega
  A_t} \left( \frac{\hbar^2}{m} \right)^2 \sum_{\substack {s_1,s_2 \\ \mu_1,\mu_2}}
\int \limits_{BZ} \frac{dk}{2\pi} \left| \mathcal{M}_{s_1s_2}^{\mu_1\mu_2} (k)
\right|^2\nonumber\\
&\times\frac{f[E_{s_1,\mu_1}(k)] -
  f[E_{s_2,\mu_2}(k)]}{E_{s_2,\mu_2}(k)- E_{s_1,\mu_1}(k) - \hbar
  \omega + i\mathit{\Gamma}}\nonumber\\
&\times \frac{1}{E_{s_2,\mu_2}(k) - E_{s_1,\mu_1}(k)}\bigg],
\end{align}
where
$ \mathcal{M}_{s_1 s_2}^{ \mu_1\mu_2} (k) = \langle s_2,\mu_2,k|
\mathbf{n} \cdot \nabla | s_1,\mu_1,k \rangle $ is the optical matrix
element corresponding to a transition from an initial state
$(s_1,\mu_1)$ to a final state $(s_2,\mu_2$) \cite{Saito2004},
$A_t = \pi (d_t/2)^2 $ is the cross section area of a SWNT, and
$f[E(k)]$ is Fermi-Dirac distribution function.  The electron wave
function $|s,\mu,k \rangle$ is related with the subband energy
$E_{s,\mu}(k)$, where $s=c$ ($s=v$) for a conduction (valence) subband
and $\mu$ is the index for the cutting line, which represents the 1D
Brillouin zone (BZ) of the SWNT~\cite{saito1998physical} with the
electron wave vector $k$.  The cutting lines are plotted in the 2D BZ
of graphene with index $\mu = 1, 2, \ldots, N$.  The value of $N$
depends on $(n,m)$ according to the formula $N=[2(n^2+m^2+nm)]/d_R$,
where $d_R=\gcd(2n+m,2m+n)$.

In Eq.~\eqref{epsilon}, $\mathit{\Gamma}$ is the broadening factor
that accounts for relaxation processes in optical transitions
resulting in finite lifetime $\tau$ of the electron state.  Here we
simply assume that $\mathit{\Gamma}$ does not depend on $\omega$ or
$E_F$ but is constant,
$\mathit{\Gamma} = 50~\mathrm{meV}$~\cite{Hertel2000}.  The numerical
integration over $k$ is implemented by the left Riemann sums
approximation, where the step $dk$ is chosen to reach an accuracy
$\Delta \varepsilon/\varepsilon = \pm 0.01$, corresponding to
$dk = \mathit{\Gamma}/(5\hbar v_F)$, where $v_F = 10^6~\mathrm{m/s}$
is the Fermi velocity in graphene.

To obtain the energy band structure of carbon nanotubes, we adopt the
zone-folding approximation of graphene with long-range atomic
interactions up to the third nearest-neighbor transfer integrals, or
the so-called third nearest-neighbor tight-binding (3rd NNTB)
model~\cite{Reich2002, Chegel2015}.  Although this approach does not
include the curvature effect, the resulting band structure is
sufficiently accurate for SWNTs with diameter larger than
$1~\mathrm{nm}$~\cite{Popov2004}.  Note that in contrast to the
simplest tight-binding approach, the subbands within the valence and
conduction bands in the 3rd NNTB model are not further symmetric with
respect to $E=0$.  Therefore, the SWNTs properties are more sensitive
to the doping type ($n$-type or $p$-type) as usually observed in
experiments.

\subsection{Optical selection rules}
\label{subsec:opt}

Both dielectric function and optical conductivity are obtained by
taking summation of different contributions from all possible pairs of
$(s_1, \mu_1)$ and $(s_2, \mu_2)$.  Although the summation in
Eq.~\eqref{epsilon} is performed over all the cutting lines in valence
and conduction bands, only limited number of subbands gives nonzero
contribution.  The~$(s_1,\mu_1) \rightarrow (s_2,\mu_2)$ transition is
contributive when $\mathcal{M}_{s_1,s_2}^{\mu_1,\mu_2}(k)$ is nonzero
(optical selection rules) and the Pauli exclusion principle is
satisfied (the difference of Fermi-Dirac distributions in
Eq.~\eqref{epsilon} is nonzero).  The concept of optical selection
rules for SWNTs was originally discussed by Ajiki and
Ando~\cite{Ajiki94}, who formulated the optical matrix elements by
current-density operator.  They proved that the allowed transitions
are always vertical ($k_1 = k_2$) and the cutting line index should be
conserved for parallel polarization ($\mu_1 = \mu_2$).  On the other
hand, the optical transition for perpendicular polarization occurs
within nearest neighbor cutting lines, $\mu_2 = \mu_1 \pm 1$.

For the sake of completeness, we rederive the optical selection rules
within the dipole approximation.  For parallel polarization, the
optical matrix elements are
\begin{align}
\label{parallel_matr}
  \mathcal{M}_{s_1 s_2}^{\mu_1 \mu_2}(k_1,k_2) =
  &  \sum_{\ell, \ell'=A,B}
     C^{s_2*}_{k_2\mu_2 \ell'} C^{s_1}_{k_1\mu_1 \ell} \nonumber  \\ 
  & \times \delta(k_1 - k_2) \delta(\mu_1 - \mu_2) \nonumber  \\ 
  & \times \sum_{j}  \mathbf{n}_{||} 
    \cdot \langle j,\ell'| \nabla|0,\ell \rangle
    e^{-i \mathbf{k}_2 \cdot \mathbf{R}(j)} ,
\end{align}
and for perpendicular polarization we obtain
\begin{align}
\label{perp_matr}
  \mathcal{M}_{s_1 s_2}^{\mu_1 \mu_2}(k_1,k_2) =
  &  \sum_{\ell, \ell'=A,B} C^{s_2*}_{k_2\mu_2 \ell'}  
    C^{s_1}_{k_1\mu_1 \ell} \delta(k_1 - k_2)  \nonumber \\
  & \times \frac{1}{2} \left( \delta(\mu_1 - \mu_2 -1)
    + \delta(\mu_1 - \mu_2 +1) \right) \nonumber \\ 
  & \times \sum_{j}  \mathbf{n}_\perp \cdot \langle j,\ell'|
    \nabla|0,\ell \rangle 
    e^{-i \mathbf{k}_2 \cdot \mathbf{R}(j)} .
\end{align}
The detailed derivation for
Eqs.~\eqref{parallel_matr}~and~\eqref{perp_matr}, as well as the
meaning of each variable in their right-hand sides, are given in
Appendix~\ref{app:selection}.  It should be noted that the results of
optical selection rules are the same either by considering dipole
approximation or current-density operator~\cite{Ajiki94}.

\begin{figure}[t!]
  \centering \includegraphics[clip,width=8cm]{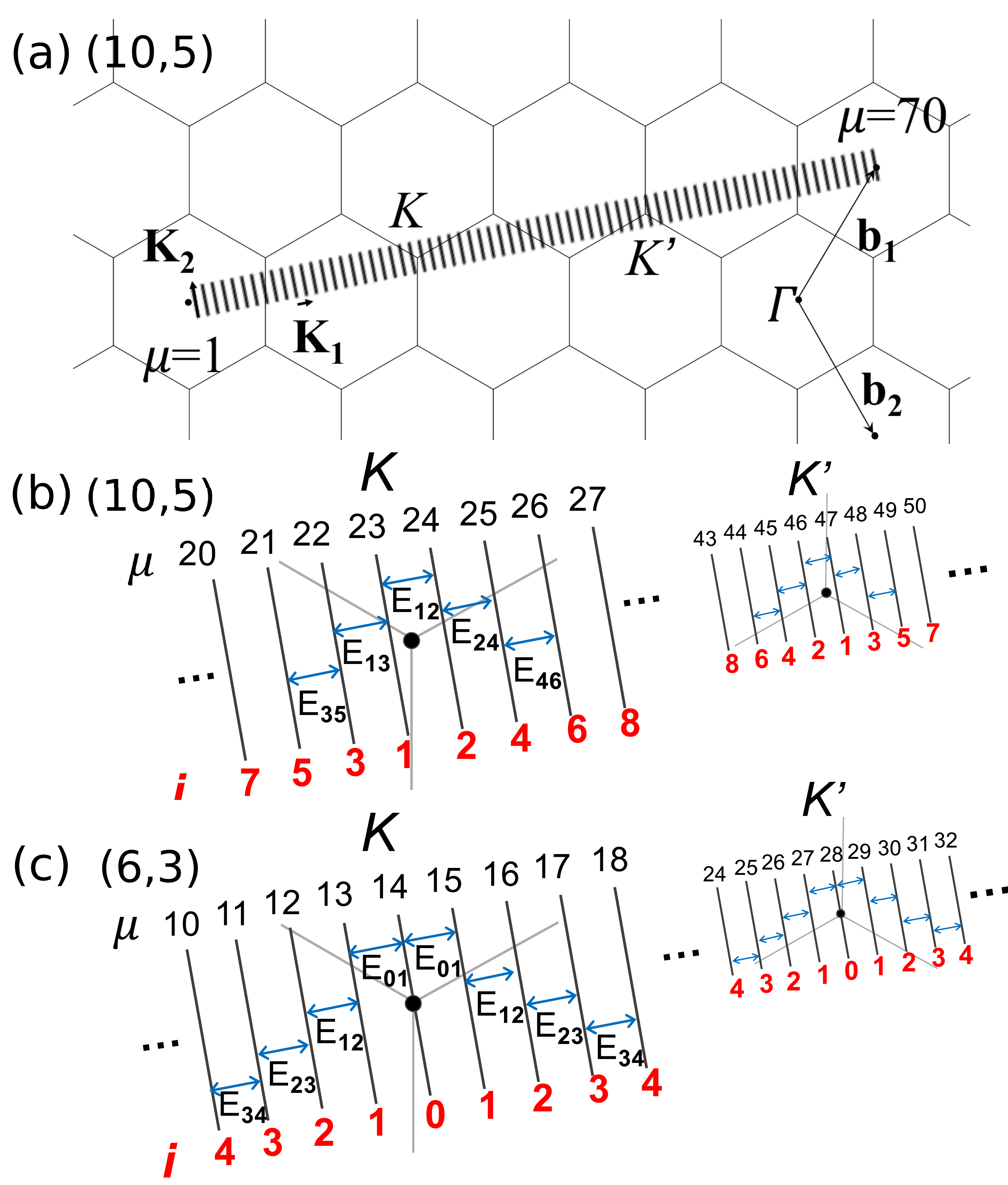}
  \caption{\label{fig:2}(a) Cutting lines of the $(10,5)$ SWNT on
    hexagonal 2D BZ, where $N = 70$.  We also show a closer look at
    cutting lines around $K$ and $K^\prime$ points for (b) $(10,5)$
    semiconducting SWNT and (c) $(6,3)$ metallic SWNT.  In (b) and
    (c), two approaches are demonstrated to label the energy bands:
    with the cutting line index $\mu$ (upper) and with optical
    transition index $i$ (lower).  Intersubband transitions for
    perpendicular polarization are shown by arrows.}
\end{figure}

When we discuss the plasma oscillations in the electron gas, all
charges are considered equivalent and contributing to the collective
motion.  However, it is not the case for SWNTs, in which the
electronic states consist of $N$ subbands in both valence and
conduction bands.  The calculated plasmonic excitations in nanotubes
show that the plasmon peak is dominated by a particular
$(s_1,\mu_1) \rightarrow (s_2,\mu_2)$ transition.  With this regard,
and also for clarity in presenting our results, let us introduce a
more convenient notation for the plasmonic transition that can be used
generally for all ($n, m$) SWNTs.  Here our target is to assign
one-to-one correspondence between the
$(s_1,\mu_1) \rightarrow (s_2,\mu_2)$ transition and the
\emph{intersubband} transition energy $E_{ij}$, similar to the
notation adopted for the \emph{interband} optical transitions $E_{ii}$
\cite{Saito2000,samsonidze2003concept}.  The case of $s_1 \ne s_2$ is
the \emph{interband} transition, while the case of $s_1 = s_2$ (with
$\mu_1 \ne \mu_2$) is the \emph{intersubband} transition.  The
condition of $s_1 = s_2$ means that we consider the intersubband
transition within the conduction (or valence) band.  Therefore,
instead of using the cutting line index $\mu$, which strongly depends
on the SWNT structure, we will label the cutting line by integers $i$
starting from the cutting line closest to the $K$ point as shown in
Figs.~\ref{fig:2}(b) and \ref{fig:2}(c) for $(10,5)$ semiconducting
and $(6,3)$ metallic SWNTs, respectively.  It is possible to
analytically obtain the new cutting line indices (optical transition
indices) around the $K$ and $K^\prime$ points~\cite{Saito2005}.  Then,
the transitions can be enumerated according to the distance of the
corresponding cutting line from the $K$ or $K^\prime$ points
[Fig. \ref{fig:2}(b)], such as $E_{12}, E_{13}, E_{24}, E_{35}$ and
$E_{46}$ for a semiconducting SWNT.  In the case of metallic SWNT
[Fig. \ref{fig:2}(c)], by excluding the trigonal warping
effect~\cite{Saito2000}, we can obtain transitions such as
$E_{01}, E_{12}$, $E_{23}$, and so on, either going to the right or
left direction away from the $K$ (or $K^\prime$) point.

\begin{figure}[t!]
  \centering \includegraphics[clip,width=8cm]{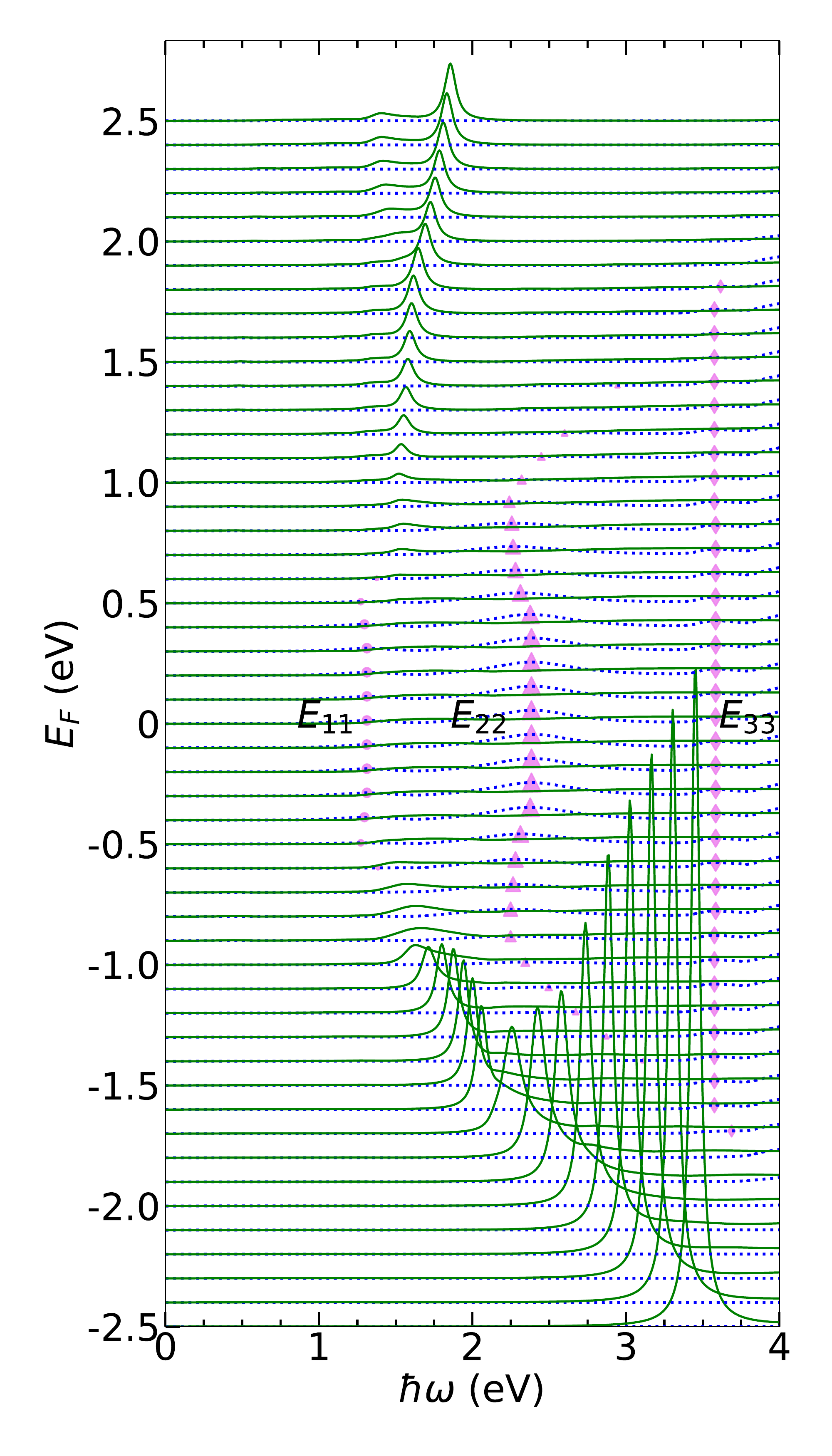}
  \caption{\label{fig:3} Doping-induced evolution of optical
    absorption spectra in a (10,5) SWNT.  Solid (dotted) lines
    represent perpendicular (parallel) polarization of light.
    Circles, triangles, and diamonds are a guide for eyes to trace the
    $E_{11}$, $E_{22}$, and $E_{33}$ transition peaks in the case of
    parallel polarization.  The absorption peaks in the case of
    parallel polarization are not due to intersubband plasmons, while
    the peaks in the case of perpendicular polarization are caused by
    intersubband plasmons, as discussed in the main text.}
\end{figure}

\section{Results and discussion}
\label{sec:res}
\subsection{Absorption spectra of doped SWNT}
\label{sec:absspec}

Let us firstly discuss the absorption spectra of doped SWNT for a
particular $(n,m)$.  In Fig.~\ref{fig:3} we plot
$\mathrm{Re} (\sigma/\varepsilon)$ of the $(10,5)$ SWNT as a function
of photon energy $\hbar \omega$ for parallel and perpendicular
polarization.  Many spectra are plotted for different Fermi energies
$E_F$ from $-2.5$ to $2.5$ eV.  For $|E_F| < 0.5$~eV, since the first
energy subband of conduction (valence) band is not occupied, we can
observe interband transitions of all $E_{ii}$'s with $i \in \{1,2,3\}$
for the transitions between the valence and conduction bands.  When we
increase $|E_F|$ more than $0.5$~eV, the $E_{ii}$ peaks start to
disappear from $E_{11}$ to $E_{33}$ because the $i$th subband in the
conduction (valence) band begins to be occupied (unoccupied) for
$i=1, 2$, and so on.  The position of $E_{ii}$ peaks (circles,
triangles and diamonds for $E_{11}$, $E_{22}$ and $E_{33}$,
respectively) is redshifted by increasing doping and then blueshifted
before disappearing.  The redshift of $E_{ii}$ occurs because of the
depolarization correction, which decreases with doping, whereas the
blueshift attests the parabolic shape of the subbands.  The
depolarization correction can be seen as the inclusion of Coulomb
interaction between electrons in the calculation of optical absorption
[$\mathrm{Re} (\sigma/ \varepsilon)$], since
$\varepsilon(\omega)=1+iv_q\sigma(\omega)q^2/(e^2\omega)$, where
$v_q=2\pi e^2/q$ is the Coulomb potential and $q=2/d_t$. Hence the
dielectric function can be expressed as in
Eq.~(\ref{conductivity}). Without the inclusion of Coulomb
interaction, the position of $E_{ii}$ absorption peaks is constant by
doping, not redshifted.  Although we do not include the excitonic
effect for simplicity, the presence of redshift in the $E_{ii}$ peaks
in our calculation is consistent with the previous work by Sasaki and
Tokura~\cite{sasaki18-plasmonpol}.  It should be noted that, by the
exclusion of excitonic effect, for $d_t=2$~nm, the deviation of the
peak positions (defined as maxima of $\mathrm{Re} [\sigma(\omega)]$)
is still less than $10$\% in comparison with the exciton Kataura
plot~\cite{Jiang2007}.

While the $i$th subband is being occupied with electrons (or holes),
the value of $E_{ii}$ increases because the single-particle
excitations occur only for the restricted $k$-regions, which are far
from $k_{ii}$~\cite{Saito2000, Saito2005, Jiang2004}, where the
interband energy distance is larger.  When the subband is partially
occupied, a new peak for perpendicular polarization appears.  We
expect that such a peak is related with intersubband plasmon
excitations for several reasons: (1)~$\mathrm{Re}(\varepsilon)$ has a
zero point close to the peak position, (2) the peak position is
different from the single-particle intersubband $i \rightarrow j$
transition, (3) the peak intensity strongly depends on Fermi energy
and continuously increases even when the subbands are almost occupied
and part of transitions is blocked, and (4) the blueshift with
increasing the Fermi energy is opposite to the redshift for the
single-particle excitation \cite{Igarashi2015}.  For highly positive
doping $E_F > 1.9$~eV, the second smaller peak is observed around
$1.4$~eV as shown in Fig.~3. This peak is another type of plasmon,
which differs from the first one at $1.5-1.8$~eV by the dominant
contributions (see the more detailed discussion in Appendix
\ref{plasm}).  Hereafter, we focus our attention to the first, main
plasmon peak, since this one should easily be observed in experiments.
The Fermi-energy dependent optical absorption shown in
Fig.~\ref{fig:3} is consistent with that previously discussed by
Sasaki and Tokura \cite{sasaki18-plasmonpol} for the armchair
$(10,10)$ and zigzag $(16,0)$ SWNTs.  However, the present result
shows additional plasmon peaks (Appendix \ref{plasm}) and different
doping-type dependence (for $E_F>0$ and $E_F<0$), which appears by
introducing more accurate energy band calculation.

\begin{figure}[t!]
  \centering \includegraphics[clip,width=8cm]{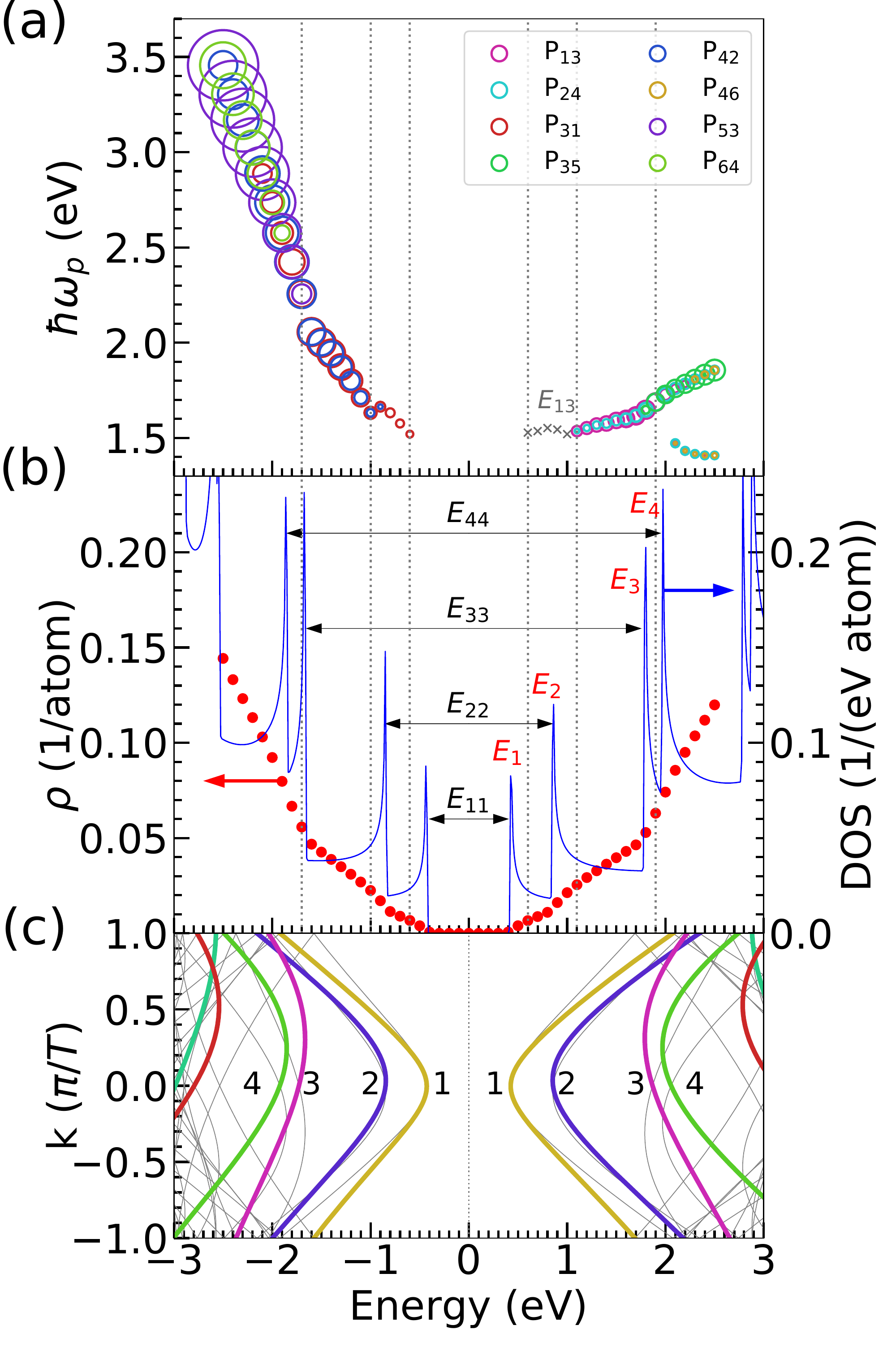}
  \caption{\label{fig:4} (a) Plasmon frequency as a function of Fermi
    energy for the $(10,5)$ SWNT. The radius of the circles
    corresponds to the intensity of the P$_{ij}$ peak.  Note that the
    weak peaks for $0.6<E_F<1.1$~eV are not plasmonic, but related to
    $E_{13}$ absorption. (b) Density of states (solid line) and charge
    density (points) for $(10,5)$ SWNT as a function of energy.
    Dotted vertical lines indicate the positions of kinks for plasmon
    frequency.  (c) Energy band structure for the $(10,5)$
    SWNT. Colored bold lines correspond to the subbands coming from
    the cutting lines nearest to the $K$ point. Thin solid lines
    correspond to the subbands from the other cutting lines in the
    presented energy range.  }
\end{figure}

\subsection{Plasmon excitation in SWNT}

In Fig.~\ref{fig:4}(a) we plot the absorption peak position in the
case of perpendicular polarization for the $(10,5)$ SWNT as a function
of $E_F$.  The intensity of each peak is represented by the circle
diameter.  We attribute the peak as the plasmon peak and denote its
frequency as $\omega_p$ when $\text{Re}[\varepsilon(\omega_0)]=0$ and
$\omega_0$ is close ($\le 20$ meV) to $\omega_p$.  Each point in
Fig.~\ref{fig:4}(a) consists of several circles which correspond to
different contributions from the transition of the cutting line pair
$i \rightarrow j$ measured from the $K$ point.  We denote the dominant
$i \rightarrow j$ contribution as P$_{ij}$, where the threshold for
dominant contribution was chosen as $10$\% of maximum contribution for
each peak.  Here we omit the valence and conduction band indices
($s_1,s_2$) since the dominant transition is the intersubband
transition, $s_1=s_2$.  One can clearly observe the kink shape of the
function, as well as the existence of the second plasmon branch at
lower frequencies for $E_F>2$~eV (see Appendix \ref{plasm} for
details).

In Fig.~\ref{fig:4}(b) we display the density of states (DOS) and
charge density as a function of Fermi energy for the $(10,5)$
nanotube.  The charge density for electrons at $E_F>0$ is given by
$\rho(E_F)= \int_0^\infty D(E) f(E) dE$, where $D(E)$ is the DOS.  For
holes at $E_F<0$ we modify the charge density formula by replacing the
distribution function $f(E)$ with $1 - f(E)$.  In Fig.~\ref{fig:4}(c)
we show energy dispersion $E_{s,\mu}(k)$, where the energy subbands
are labeled according to the approach discussed in
Sec.~\ref{subsec:opt}.  The kink positions for the plasmon energy and
the charge density $\rho(E_F)$ are shown to be consistent to each
other [see grey dotted lines in Fig.~\ref{fig:4}(b)].  In the
three-dimensional (3D) Drude model, the plasmon frequency is known to
be proportional to the square root of charge density
($\omega_p^{3\mathrm{D}} \propto \sqrt \rho$ ).  For carbon nanotubes,
the Fermi energy dependence was predicted to be consistent with 2D
graphene result
($\omega_p^{2\mathrm{D}} \propto \sqrt
E_F$)~\cite{Abajo2014}. However, we see from Figs.~\ref{fig:4}(a)
and~\ref{fig:4}(b) that the plasmon frequency is a function of
$\rho(E_F)$, which in case of carbon nanotubes is the sum
$\sum_{E_{ii}<E_F} \sqrt{E_F - E_{ii}} $.

\begin{figure}[t!]
  \centering \includegraphics[clip,width=8cm]{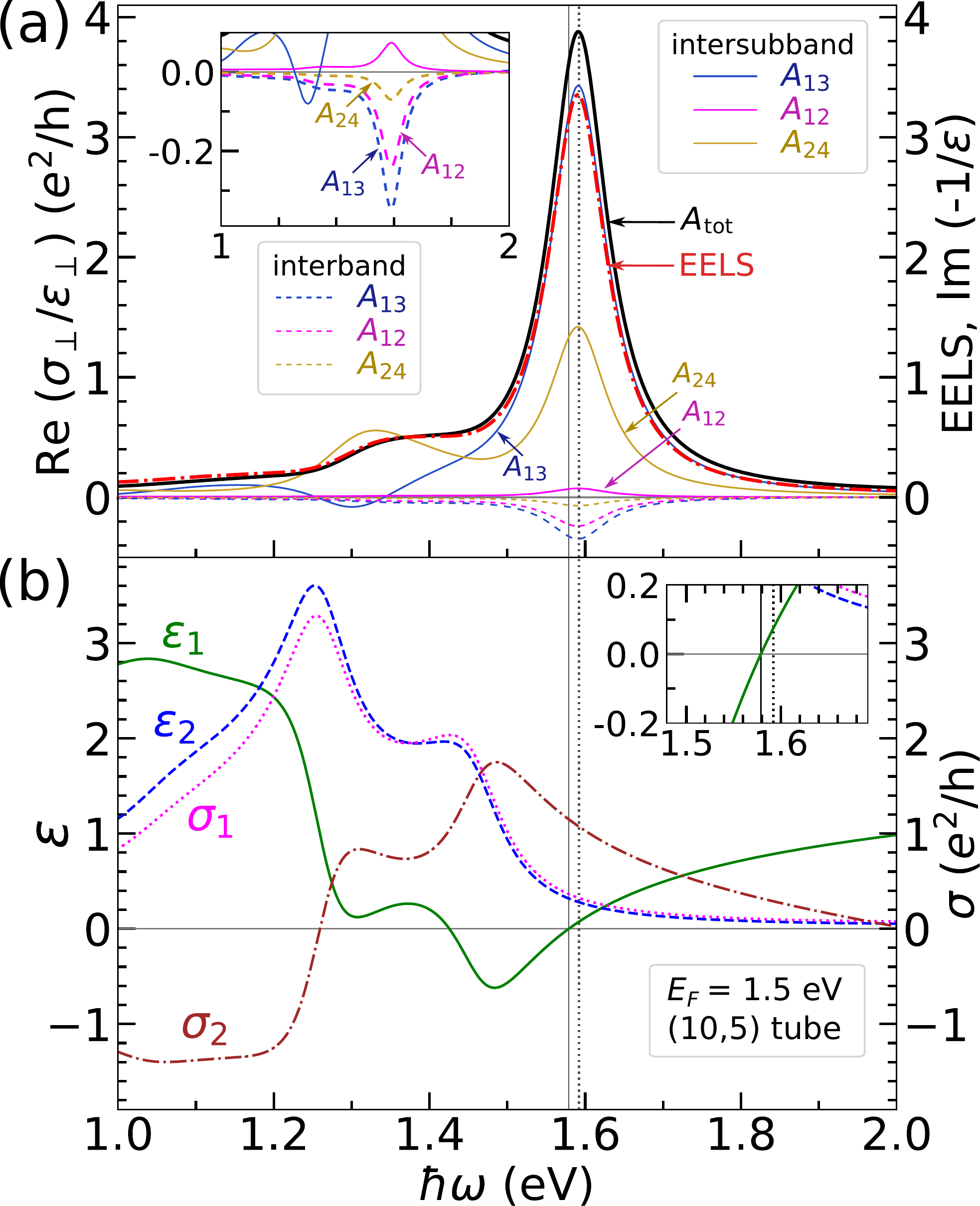}
  \caption{\label{fig:5} (a) Absorption spectra for a doped $(10,5)$
    SWNT with $E_F=1.5$ eV.  Black bold solid line represents the
    total absorption $A_\mathrm{tot}$, considering both the
    intersubband and interband transitions.  Colored solid lines
    correspond to the dominant $A_{13}$, $A_{24}$, and $A_{12}$
    \emph{intersubband} contributions.  The EELS spectrum,
    $\mathrm{Im} (-1/\varepsilon)$, is plotted with red dash-dotted
    line.  Colored dashed lines lines correspond to the interband
    absorptions with the same transition indices as the intersubband
    counterparts.  Inset depicts the enlarged region for the interband
    peaks, which are about one order-of-magnitude smaller than the
    intersubband peaks.  (b) Real ($\varepsilon_1$) and imaginary
    ($\varepsilon_2$) parts of dielectric function along with
    conductivity ($\sigma_1$ and $\sigma_2$) for $(10,5)$ doped SWNT
    with $E_F=1.5$ eV.  Solid (dotted) vertical line corresponds to
    $\mathrm{Re}(\varepsilon) = 0$ [$\rm{max}(A_\mathrm{tot})$].}
\end{figure}

The kink in $\rho(E_F)$ appears when $E_F$ passes through the next van
Hove singularity ($E_{i}$) as shown in Fig.~\ref{fig:4}(b), which is
followed by the Pauli blockade of the $i$th subband and change in the
dominant contribution to the plasmon from P$_{ij}$ to another
P$_{i'j'}$, where $i'>i$ and $j'>j$ for $E_F>0$ ($i'<i$ and $j'<j$ for
$E_F<0$). As seen from Fig.~\ref{fig:4}(a), the first dominant
contribution is P$_{13}$ (P$_{31}$), the second dominant contribution
after the first kink is P$_{24}$ (P$_{42}$), the third contribution
after the second kink is P$_{35}$ (P$_{53}$) for $E_F>0$ ($E_F<0$).
The plasmon intensity [radius of circle in Fig.~\ref{fig:4}(a)]
increases with increasing the Fermi energy and inceasing $\rho (E_F)$.

The asymmetry of plasmon peak intensity with respect to the $n$-type
and $p$-type doping is consistent with asymmetric nature of
$\rho (E_F)$ for $E_F>0$ and $E_F<0$. The minimum plasmon frequency as
well as the Fermi energy at which the plasmon is excited basically
depend on the energy band structure. For example, in
Fig.~\ref{fig:4}(a), the asymmetry in the values of $E_{13}$ within
valence and conduction bands influences the starting plasmon frequency
($\hbar \omega_p = 1.52$~eV for the valence band and
$\hbar \omega_p = 1.54$~eV for the conduction band). Meanwhile, the
number of subbands under or above the Fermi level within the valence
or conduction band is essential for accumulating negative contribution
to dielectric function in order to observe
$\mathrm{Re}(\varepsilon) = 0$. Therefore, the interplay between the
intersubband transitions determines the asymmetric nature of the
plasmon peak intensity in the $n$-type and $p$-type doping.  Note that
at $E_F = 0$~eV both real and imaginary parts of $\varepsilon(\omega)$
are positive in the energy range of $0-4$~eV. In the case of p-doped
(10,5) SWNT, the plasmon starts to appear at $E_F=0.6$~eV, after the
1st subband becomes partially unoccupied, in which the condition of
$\mathrm{Re}(\varepsilon) = 0$ is already satisfied. In the case of
n-doping, the first small peak appears at $E_F=1.1$~eV.  However,
since $\mathrm{Re}(\varepsilon) \ne 0$, this peak is still not a
plasmon, but is a single-particle intersubband transition
$1 \rightarrow 3$. It is observed when the 1st subband is partially
occupied and when the depolarization effect, which was completely
suppressing absorption before, is relaxed. The true plasmon peak
appears at $E_F=1.1$~eV, which corresponds to the 2nd subband
partially occupied. Thus, the condition to observe the plasmon in SWNT
for perpendicularly polarized light is to shift the Fermi level up
higher than the bottom of the 2nd subband in conduction band
\cite{sasaki16-interplasmon, yanagi18-isbp}, or down lower than the
top of the 1st subband in valence band.

In Fig.~\ref{fig:5}(a), we plot intersubband and interband absorption
spectra in case of perpendicular polarization for $(10,5)$ SWNT and
$E_F=1.5$~eV. We define the absorption associated with the
$i \rightarrow j$ transition as
$A_{ij} = \mathrm{Re} (\sigma_\perp^{ij}/\varepsilon_\perp)$, where
$\sigma_\perp^{ij}$ is
\begin{align}
\label{ij_absorpt}
\sigma_\perp^{ij} = & \frac{16}{d_t} \frac{e^2}{h} \left(
\frac{\hbar^2}{m} \right)^2 \int \limits_{-\pi/T}^{\pi/T}
\frac{dk}{2\pi} |\mathcal{M}^\perp_{ij} (k)|^2 \nonumber \\ 
& \times \frac{f(E_{i}(k)) - f(E_{j}(k)) }{E_{j}(k)- E_{i}(k) - \hbar \omega + i
  \mathit{\Gamma}} \cdot \frac{1}{E_j (k)- E_i (k)}.
\end{align} 
For $E_F>0$, when we consider the interband transitions, the $i$th and
the $j$th subbands come from the valence and conduction band
respectively.  On the other hand, for the intersubband transitions,
both subbands lie within the conduction band.  The total absorption
$A_{\mathrm{tot}}$ in Fig.~\ref{fig:5}(a) is contributed from all the
interband and intersubband transitions.  We see that the peak position
and line shape of the absorption spectrum are consistent with those of
EELS spectrum, which is given by $\mathrm{Im}(-1/\varepsilon)$.

As we already mentioned above, both optical conductivity and
dielectric function are superpositions of contributions
($\sigma_{ij}$, $\varepsilon_{ij}$) from different transitions between
the $i \rightarrow j$ subbands. To calculate absorption from the
$i \rightarrow j$ transition $A_{ij}$, we take only the corresponding
term from the conductivity $\sigma_{ij}$, while the dielectric
function ($\varepsilon_\perp$) is calculated for all pairs of
interband and intersubband transitions according to
Eq.~\eqref{epsilon}.  As an example, in the case of $E_F=1.5$~eV in
Fig.~\ref{fig:4}(a) two main contributions are P$_{13}$ and P$_{24}$.
In Fig.~\ref{fig:5}(a), we see the peak value of $A_{ij}$ for
intersubband absorption (solid lines) is one order-of-magnitude larger
than that for interband absorption (dashed lines), which clearly shows
that the plasmon has an intersubband nature.  One may notice that the
same P$_{13}$ and P$_{24}$ transitions are dominant for both
intersubband and interband absorptions.  However, the contributions
have different signs and different order of magnitude.

Although the interband transitions seem to give negligible
contribution to the plasmon intensity, they affect the redshift of the
zero point for the dielectric function~\cite{sasaki18-plasmonpol}, as
shown in Fig.~\ref{fig:5}(b).  In fact, the position of the maximum in
absorption spectra (dotted vertical line) and the zero of
$\mathrm{Re}[\varepsilon(\omega)]$ (solid vertical line) are slightly
different (by
$\sim$$1$~meV).  This difference comes from
$\mathrm{Im}[\varepsilon(\omega)]$, which decreases in the proximity
of $\mathrm{Re}[\varepsilon(\omega)] =
0$, as well as
$\mathrm{Im}[\sigma(\omega)]$ [Fig. \ref{fig:5}(b)]. If the dielectric
function is a real function of
$\omega$, the zero value would give the exact position of plasmon,
which is not the case for a complex
$\varepsilon(\omega)$.  Indeed, for $\varepsilon = \varepsilon_1 + i
\varepsilon_2$ and $\sigma = \sigma_1 + i
\sigma_2$, the absorption and the energy loss-function have the
following form:
\begin{align}
  \mathrm{Im} \left( - \frac{1}{\varepsilon} \right)
  & = \frac{1} { \varepsilon_2 \left[ 1 + \left(
    \varepsilon_1/\varepsilon_2
    \right)^2\right]},\\ \mathrm{Re} \left(
  \frac{\sigma}{\varepsilon} \right)
  & = \frac{ \sigma_2 + \sigma_1
    \left(\frac{\varepsilon_1}{\varepsilon_2}\right)}{\varepsilon_2 \left[ 1 +
    \left( \varepsilon_1/\varepsilon_2 \right)^2 \right]}.
\end{align}
The maxima of $\mathrm{Im}(-1/\varepsilon)$ and $\mathrm{Re}
(\sigma/\varepsilon)$ appear close to the $\varepsilon_1 = 0$, but not
exactly at this point. The shift of the maxima strongly depends on slope
of $\varepsilon_2(\omega)$ near the zero point of $\varepsilon_1$.

\begin{figure}[t!]
  \centering \includegraphics[clip,width=8cm]{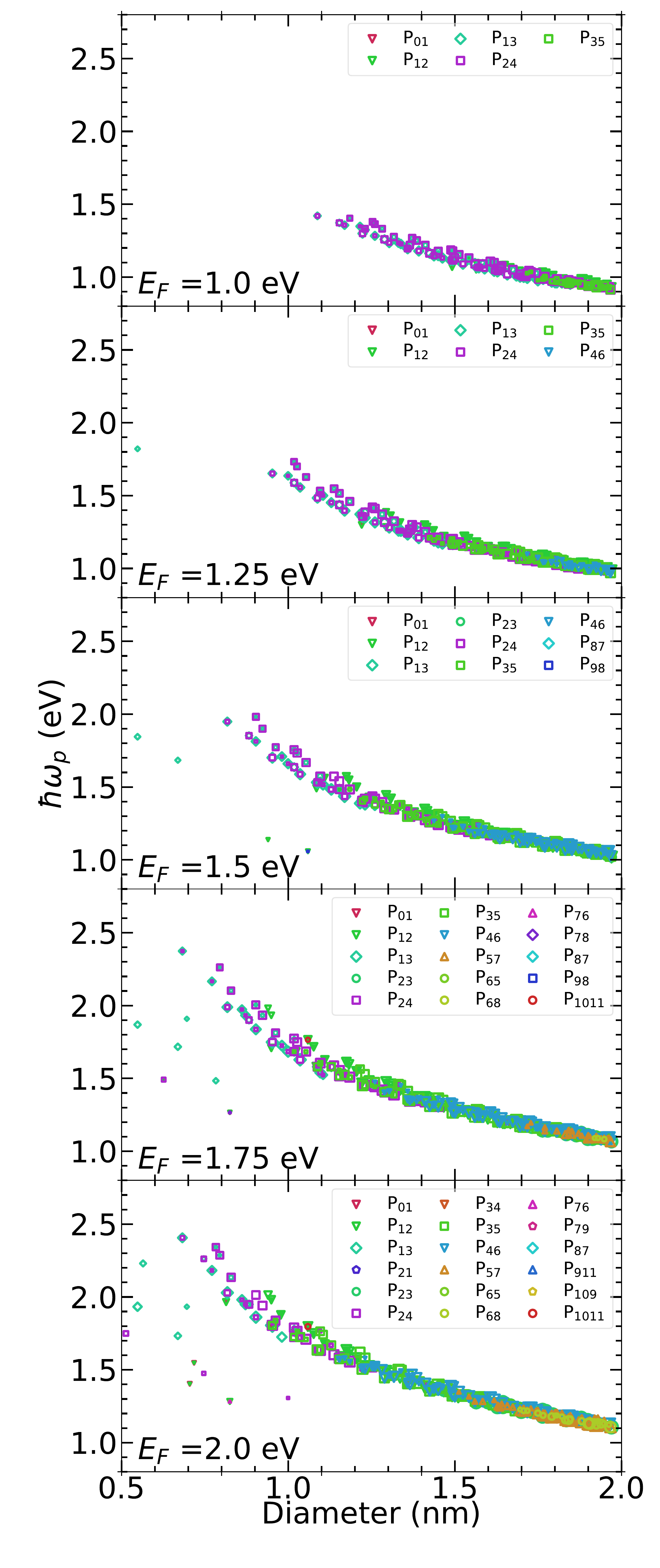}
  \caption{\label{fig:6} Intersubband plasmon frequencies (major peak)
    for SWNTs of all different chiralities $(n,m)$ with diameters from
    $0.5$ to $2~\mathrm{nm}$.  Five different Fermi energies from
    $1.0$ to $2.0$ eV are considered. The dominant contributions are
    pointed out for each plasmon (and thus each chirality) by specific
    marker types and colors.  The size of marker corresponds to the
    plasmon peak intensity~\cite{outputPlasmon}.}
\end{figure}

\subsection{Mapping of intersubband plasmons}
In Fig.~\ref{fig:6}, we plot energy of intersubband plasmon
$\hbar \omega_p$ as a function of nanotube diameter $d_t$, where
$0.5 < d_t < 2$~nm, for five Fermi energies from $E_F=1$ to 2 eV. For
$E_F = 1$ eV plasmons are observed only in tubes with $d_t > 1$
nm. With increasing $E_F$, the number of tubes which have plasmonic
excitations increases, since $E_{ii} <E_F$ ($E_{ii} \propto 1/d_t$) is
satisfied for a large $E_F$ even for smaller $d_t$ nanotubes.  Plasmon
energies $\hbar \omega_p$, as well as their spreading for fixed $d_t$
and $E_F$, are increasing with decreasing diameter.  This indicates
the presence of chirality dependence, which was neglected in the
previous works~\cite{sasaki16-interplasmon, sasaki18-plasmonpol,
  Abajo2014}.  We see that the dominant contributions for smaller
diameters and higher Fermi energies come from the cutting line pairs,
which are close to the $K$ point.  Therefore, the family spread due to
the curvature effect is inherited by plasmon frequency.  Hereafter, we
focus on the Fermi energy and diameter dependence of plasmon
frequency, since this information is useful for most experimental
studies like the Kataura plot for optical absorption
\cite{Sato2007,Nugraha2010} or Raman spectroscopy \cite{Sato2010}.
Chirality dependence of plasmon energy is a challenging point for the
present method, since the band structure calculation by adopting the
3rd NNTB model is not satisfactory to build reliable chiral angle
dependence or curvature effect~\cite{Samsonidze2004}.

\begin{figure}[t!]
  \centering \includegraphics[clip,width=8cm]{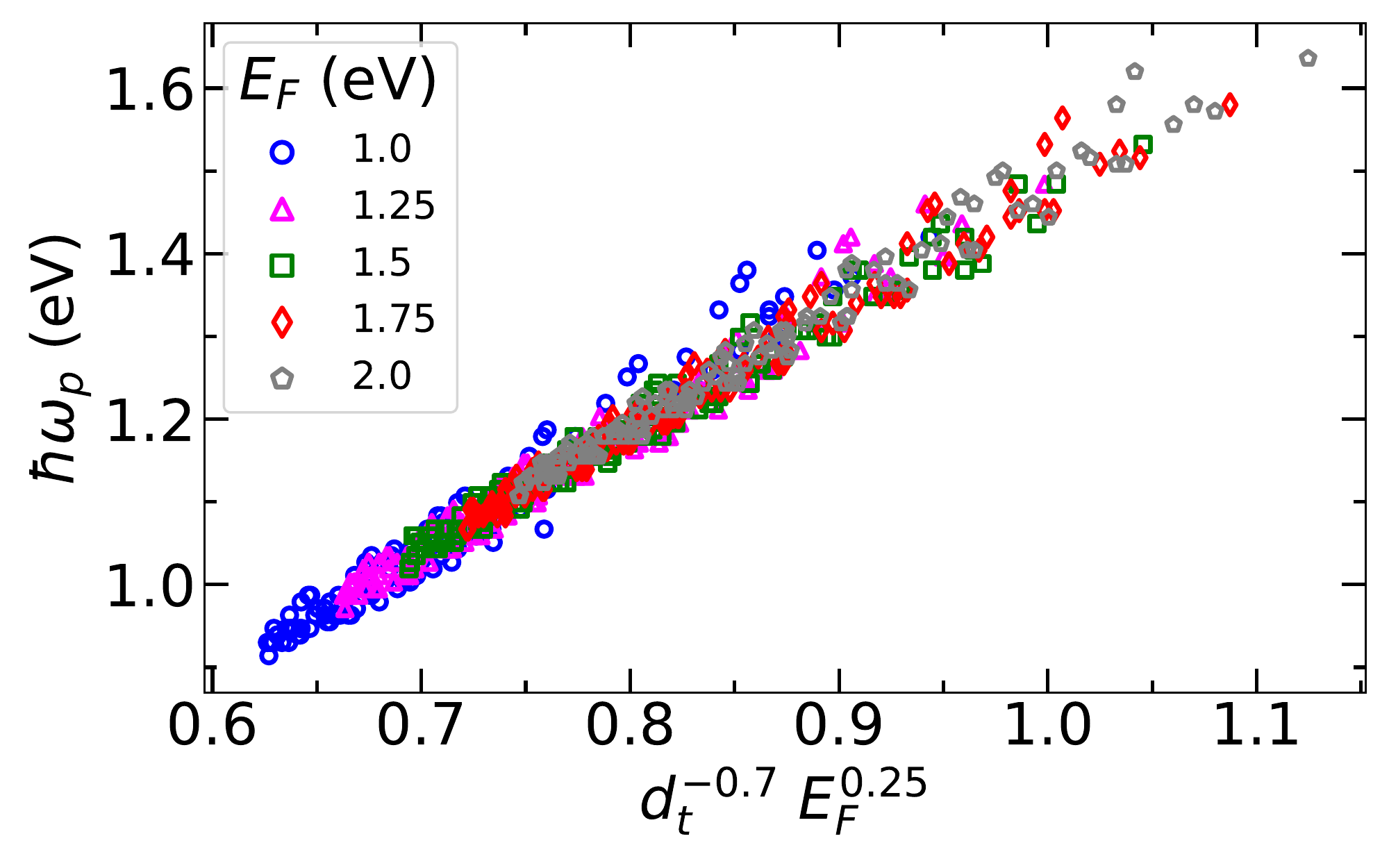}
  \caption{\label{fig:7} Fitting of the intersubband plasmon energy as
    a function of nanotube diameter $d_t$ and Fermi energy $E_F$.  We
    consider SWNTs with $1< d_t <2$ nm and only the major plasmon
    peak.}
\end{figure}

\begin{figure}[t!]
  \centering \includegraphics[clip,width=8cm]{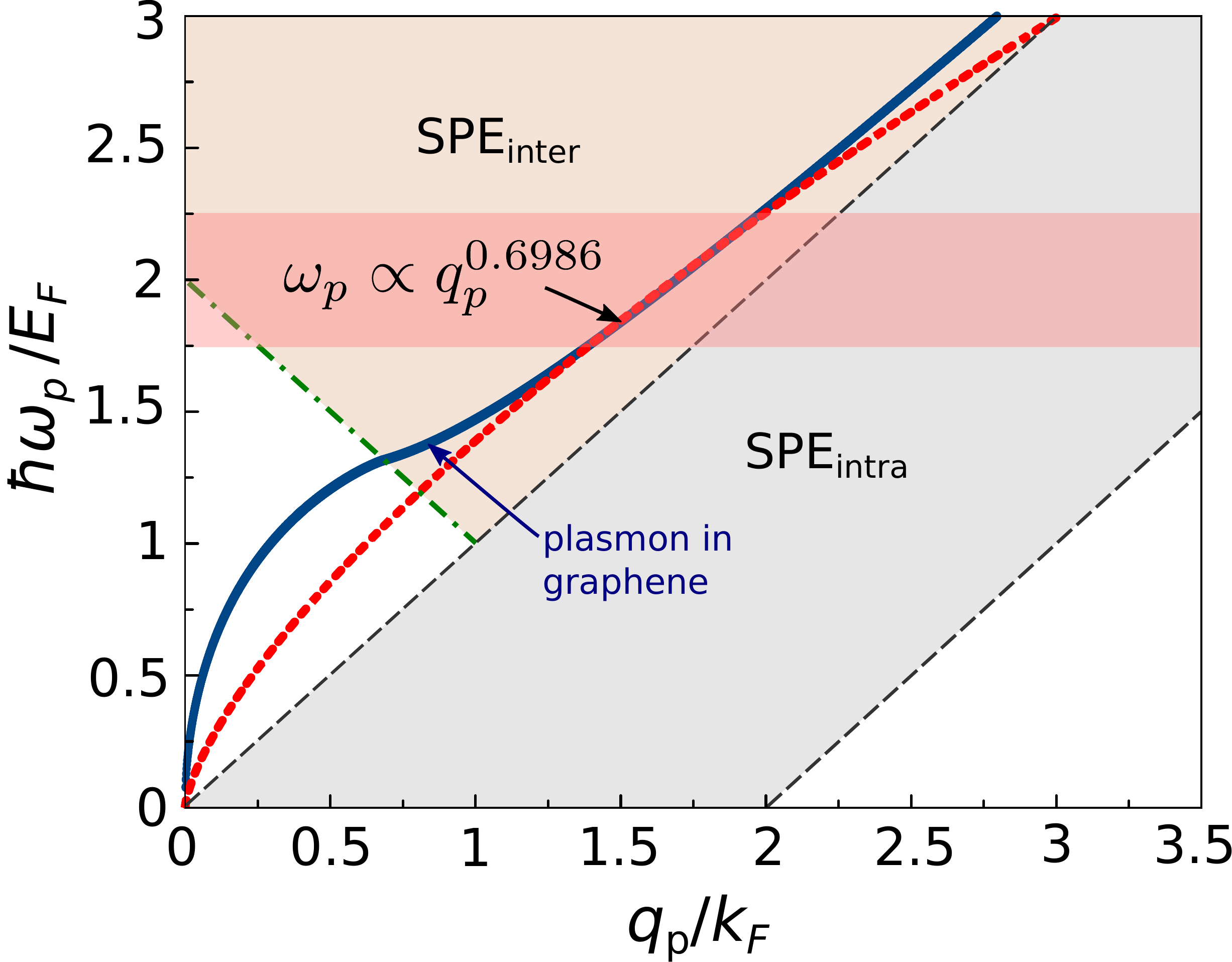}
  \caption{\label{fig:8} Fitting of the plasmon dispersion of
    graphene.  We found $\omega_p\propto q^{0.6986}$ within the
    horizontally dashed frequency range
    $(1.75~E_\mathrm{F}<\hbar\omega<2.25~E_\mathrm{F})$ that could be
    related with the intersubband plasmon excitations in SWNTs.  The
    other colored dashed areas correspond to the regime where the
    interband and intraband single-particle excitations occur in
    graphene, denoted by SPE$_{\text{inter}}$ and
    SPE$_{\text{intra}}$, respectively~\cite{hwang-graphene}.}
\end{figure}

We numerically fit the diameter and the Fermi energy dependence with
power law, as shown in Fig.~\ref{fig:7}.  The result is:
\begin{equation}
\label{plasmon_freq_fit_d}
\hbar \omega_p = (1.49 \pm 0.004) \frac{E_F^{0.25 \pm
    0.003}}{d_t^{0.69 \pm 0.005}}~\mathrm{eV}.
\end{equation}
The $d_t$ (in nm) and $E_F$ (in eV) dependence in Fig.~\ref{fig:7} can
be understood from the dispersion of plasmon in graphene, which is
shown in Fig.~\ref{fig:8}~\cite{hwang-graphene}.  The intersubband
plasmons in doped SWNTs, which are nothing but the azimuthal
plasmons~\cite{sasaki16-interplasmon}, can be considered as the
plasmons in the rolled graphene sheet, where we have the oscillations
of charge around the nanotube axis.  Rolling of graphene into SWNT
results in the quantization of plasmon wave vector ($\mathbf{q_p}$)
following the reciprocal lattice vector
$\mathbf{K}_1$~\cite{saito1998physical} in the SWNT since we consider
the transitions of electron between different cutting lines.  The
magnitude of the reciprocal lattice vector is inversely proportional
to the diameter, i.e., $|\mathbf{K}_1| = 2/d_t$, similar to the wave
vector of the electron along the circumferential direction
$(q\propto d_t^{-1})$.  From Fig.~\ref{fig:8}, we can see that the
$\sqrt{q_p}$ dependence does not always hold for plasmon in
graphene. The plasmon dispersion becomes almost linear to $q_p$ as it
enters the interband single-particle excitation (SPE$_{\text{inter}}$)
regime~\cite{hwang-graphene}.  At the colored frequency range
$(1.75~E_F<\hbar\omega_p<2.25~E_F)$ in Fig.~\ref{fig:8} we fit the
dispersion, where we get $\omega_p\propto q_p^{0.6986}$.  Therefore,
we expect $\omega_p\propto d_t^{-0.7}$ for the plasmon frequency of
SWNT, which confirms our finding in Eq.~\eqref{plasmon_freq_fit_d}. It
is noted that the $\omega_p \propto q_p^{0.7}$ of graphene's plasmon
is at relatively higher frequency range compared with the obtained
plasmon frequency range for SWNTs as shown in Fig.~\ref{fig:6}. This
is owing to the fact, that in SWNT the lower limit of photon energy
for single particle excitation (the dash-dotted line in
Fig.~\ref{fig:8}) would be smaller compared with the case of graphene
due to the possible intersubband excitation of electron within the
conduction band of SWNT. This lowering of energy limit for starting
single particle excitation by intersubband transition
(SPE$_\text{inter}$) shifts the ``almost'' linear dispersion of
plasmon in graphene to lower frequency range, too.  Thus the fitting
to ``the almost linear dispersion'' is justified.

The Fermi energy dependence of azimuthal plasmon in SWNT given by
Eq.~\eqref{plasmon_freq_fit_d} can be also understood from the
dispersion of plasmon in graphene shown in Fig.~\ref{fig:8}.  Since
the dispersion of plasmon in graphene is normalized to the Fermi
energy as shown in Fig.~\ref{fig:8}, we can obtain the following
equation:
\begin{equation}
  \label{plasmonef}
  \hbar \omega_p = \left(\frac{q_p}{k_F}\right)^{\alpha}E_F
  =(q_p \hbar v_F)^{\alpha}E_F^{1-\alpha},
\end{equation}
where we use linear energy band of graphene, $E_F=\hbar v_F k_F$.
Since $\omega_p\propto q_p^{0.6986}$, we expect the Fermi energy
dependence to be $\omega_p\propto E_F^{0.3}$, which is not exact but
close to the obtained power law in Eq.~\eqref{plasmon_freq_fit_d}. The
difference with the obtained power law comes from the fact that the
electron energy bands of SWNTs are not exactly linear as in
graphene. It is noted that if we have the $\sqrt{q_p}$ dependence of
plasmon frequency in graphene, using Eq. (\ref{plasmonef}), we will
have $\omega_p\propto E_F^{0.5}$ as expected in the Drude
model~\cite{hwang-graphene,sasaki16-interplasmon, Abajo2014}.

\section{Conclusion}
\label{sec:con}

We have systematically studied intersubband plasmon excitations in
doped SWNTs as a function of diameter and the Fermi energy.  The
intersubband plasmons are excited due to the absorption of light with
linear polarization perpendicular to the nanotube axis.  The
calculated plasmon frequency $\omega_p$ scales with the SWNT diameter
$d_t$ and the Fermi energy $E_F$ as $\omega_p \propto
(E_F^{0.25}/d_t^{0.7})$, which is a direct consequence of collective
intersubband excitations of electrons in the doped SWNTs, but
\emph{not} a result of intraband transitions described by the Drude
model.  We also show that more than one branch of intersubband
plasmons occurs even in one nanotube chirality.  Our mapping of
intersubband plasmon frequency may serve as a guide for
experimentalists to search intersubband plasmons in many different
SWNTs.

\begin{acknowledgements}
D.~S. thanks Skolkovo Institute of Science and Technology for
financially supporting a three-month visit to Tohoku University for
working on most parts of this project. A.R.T.N. acknowledges the
Interdepartmental Program for Multidimensional Materials Science
Leaders in Tohoku University.  M.~S.~U. and R.~S.  acknowledge JSPS
KAKENHI Grant Nos. JP18J10199 and JP18H01810, respectively.  A.~G.~N.
acknowledges Russian Science Foundation (Project identifier:
17-19-01787).
\end{acknowledgements}

\appendix

\section{Optical matrix elements}
\label{app:selection}

The Schr$\rm \ddot{o}$dinger equation for a SWNT is given by:
\begin{equation}
\label{shrod}
H(\mathbf{r})\psi^s_{\mathbf{k}}(\mathbf{r})=E^s_{\mathbf{k}}
\psi^s_{\mathbf{k}}(\mathbf{r}),
\end{equation}
where $H(\mathbf{r})$ is the real-space Hamiltonian, $\mathbf{k}$ is
the electron wave vector, and $s=c$ ($s=v$) denotes the conduction
(valence) band.  The wave function $\psi^s_{\mathbf{k}}(r)$ can be
expanded by a linear combination of the Bloch functions
$\phi_{\mathbf{k} \ell}(r)$ as follows:
\begin{equation}
\label{wf_tot}
\psi^s_{\mathbf{k}}(\mathbf{r})=\sum_{\ell=A,B} C^s_\ell(\mathbf{k})
\phi_{\mathbf{k} \ell}(\mathbf{r}),
\end{equation}
where $C^s_\ell(\mathbf{k})$ is the coefficient for the state
$\mathbf{k}$.  The Bloch function is expressed by
\begin{equation}
\label{atom_orbit}
\phi_{\mathbf{k}\ell}(\mathbf{r})=\frac{1}{\sqrt{N}} \sum_j
e^{i\mathbf{k} \cdot \mathbf{R}(j)} \chi (\mathbf{R}(j)-\mathbf{r}_\ell-\mathbf{r}),
\end{equation}
where $\chi (\mathbf{r})$ denotes the 2$p_z$ atomic orbital,
$\mathbf{R}(j) = j_1\mathbf{a}_1 + j_2\mathbf{a}_2$ gives the position
of the $j$th unit cell (with $\mathbf{a}_1$ and $\mathbf{a}_2$ unit
vectors of hexagonal unit cell \cite{saito1998physical}),
$\mathbf{r}_\ell$ is the position of $\ell$th atom (A or B) in the
$j$th unit cell, and $N$ is the number of unit cells.  Substituting
Eq.~(\ref{wf_tot}) into Eq.~(\ref{shrod}) we obtain:
\begin{align}
\label{shrod2}
\frac{1}{\sqrt{N}}\sum_{\ell=A,B} C^s_\ell (\mathbf{k}) \sum_j 
e^{i\mathbf{k} \cdot \mathbf{R}(j)}  H(\mathbf{r}) \chi
(\mathbf{R}(j)-\mathbf{r}_\ell-\mathbf{r}) \nonumber \\ 
 = E^s_{\mathbf{k}} \frac{1}{\sqrt{N}}\sum_{\ell=A,B}
  C^s_\ell(\mathbf{k}) \sum_j e^{i\mathbf{k} \cdot \mathbf{R}(j)} 
  \chi (\mathbf{R}(j)-\mathbf{r}_\ell-\mathbf{r}) .
\end{align}

One can rewrite Eq.~(\ref{shrod2}) in a matrix form multiplying
$\chi (\mathbf{R}(\mathbf{0})-\mathbf{r}_{\ell'}-\mathbf{r})$ to
Eq.~(\ref{shrod2}):
\begin{equation}
\label{shrod_matrix}
\sum_{\ell=A,B} C^s_\ell (\mathbf{k}) H_{\mathbf{k}\ell'\ell}= 
\sum_{\ell=A,B} E^s_{\mathbf{k}} C^s_\ell (\mathbf{k}) S_{\mathbf{k}\ell'\ell},
\end{equation}
where $H_{\mathbf{k}\ell'\ell}$ and $S_{\mathbf{k}\ell'\ell}$ are $2 \times 2$  
Hamiltonian and overlap matrices respectively, defined by:
\begin{align}
\label{matr_el}
  H_{\mathbf{k}\ell'\ell} & = \sum_j e^{i\mathbf{k} \cdot \left( \mathbf{R}(j) \right)}
                            H_{\ell'\ell}(j), \\
  S_{\mathbf{k}\ell'\ell} & = \sum_j e^{i\mathbf{k}\cdot \left( \mathbf{R}(j) \right)}
                            S_{\ell'\ell}(j),
\end{align}
and
\begin{align}
\label{matr_integrals}
H_{\ell'\ell}(j) & = \int d\mathbf{r} \chi (\mathbf{R}(\mathbf{0})-
\mathbf{r}_{\ell'}-\mathbf{r})H\chi (\mathbf{R}(j) - \mathbf{r}_\ell - \mathbf{r}), \\
S_{\ell'\ell}(j) & = \int d\mathbf{r} \chi (\mathbf{R}(\mathbf{0})-
\mathbf{r}_{\ell'} - \mathbf{r}) \chi (\mathbf{R}^l(j) - \mathbf{r}_\ell - \mathbf{r}). 
\end{align}
$H_{\ell'\ell}(j), S_{\ell'\ell}(j)$ are considered up to the third nearest neighbor sites. 
Thus we come to the generalized problem for eigenvectors and eigenvalues of the form:
\begin{equation}
\label{matrix_prob}
H_{\mathbf{k}} \mathbf{C}^s_{\mathbf{k}} = E^s_{\mathbf{k}} S_{\mathbf{k}} 
\mathbf{C}^s_{\mathbf{k}}, 
\end{equation}
where $E^s_{\mathbf{k}}=\{E^v_{\mathbf{k}}, E^c_{\mathbf{k}}\}$ gives
the energy of valence and conduction subbands for particular SWNT and
vector
$\mathbf{C}^s_{\mathbf{k}} = (C^s_A(\mathbf{k}), C^s_B(\mathbf{k}))^T$
gives the coefficients for the wave function represented by
Eq.~(\ref{wf_tot}). Within the zone-folding approach
Eq.~(\ref{matrix_prob}) is solved for Hamiltonian of 2D graphene,
while the wave vector is taken as quasi-1D BZ for SWNT given by
\cite{saito1998physical}:
\begin{equation}
\label{kvector}
\mathbf{k}=k \frac{\mathbf{K_2}}{|\mathbf{K_2}|} + \mu \mathbf{K_1}, 
\quad (\mu = 1,\ldots,N;~-\frac{\pi}{T} \le k \le  \frac{\pi}{T}),
\end{equation}
where $T$ is the length of translational vector $\mathbf{T}$,
$N=[2(n^2+m^2+nm)]/d_R$ is the number of cutting lines and
$\mathbf{K}_2$ denotes one-dimensional reciprocal lattice vector
\cite{saito1998physical}.  We adopt the wave vector notation of
Eq.~(\ref{kvector}) to the single electron wave function in carbon
nanotube as bra-ket style as
$|s, \mu, k \rangle \equiv \psi^s_{\mathbf{k}}(\mathbf{r})$.

\begin{figure}[t!]
  \centering \includegraphics[clip,width=8.5cm]{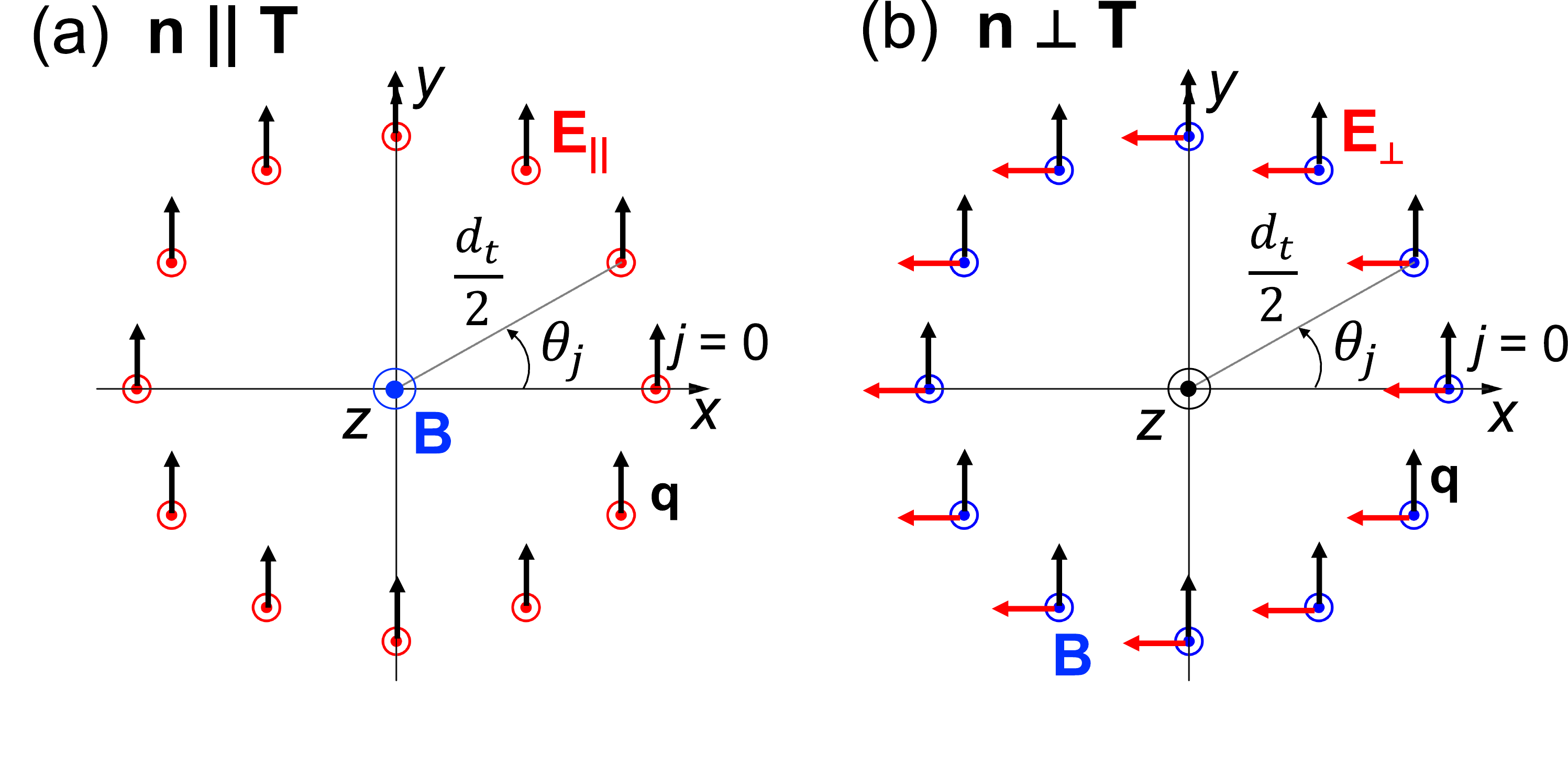}
  \caption{\label{fig:9} Projections of probe photon wave vector
    $\mathbf{q}$ and electric field $\mathbf{E}$ onto nanotube cross
    section for (a) parallel polarization and (b) perpendicular
    polarization.}
\end{figure}
The single-particle Hamiltonian in the presence of external
electromagnetic field is given by:
\begin{equation}
\label{pert_ham}
H(\mathbf{r},t) = H(\mathbf{r}) + \frac{i \hbar e}{m}
\mathbf{A}(\mathbf{r},t) \cdot \nabla,
\end{equation}
where $e>0$ is elemental charge and $m$ is the mass of electron. The
optical matrix element is given by
$ \langle s_2,\mu_2,k_2 | \frac{i \hbar e}{m} \mathbf{A_q} \cdot
\nabla | s_1, \mu_1, k_1 \rangle$, where $\mathbf{A_q}$ is Fourier
component of the vector potential
$\mathbf{A}(\mathbf{r}, t) = A_0 \mathbf{n} \cos (\mathbf{q} \cdot
\mathbf{r} - \omega t)$.  For the light propagating parallel to the
nanotube axis ($\mathbf{n} \parallel \mathbf{T}$)
[Fig.~\ref{fig:9}~(a)], $\mathbf{A_q}$ in the $j$th unit cell can be
expressed as \cite{Sato2017}
\begin{align}
\label{vect_par}
\mathbf{A_q}^{||} (\mathbf{R}(j) ) = & A _0  \mathbf{n_{||}} 
e^{i \mathbf{q} \cdot \mathbf{R}(j)}  \nonumber \\ 
= & A _0  \mathbf{n_{||}} \left( 1 + i q \frac{d_t}{2} \sin \theta_j \right),
\end{align}
In the case of perpendicularly polarized light
($\mathbf{n} \perp \mathbf{T}$) [Fig.~\ref{fig:9}(b)], $\mathbf{A_q}$
is expressed as
\begin{align}
\label{vect_perp}
\mathbf{A_q}^{\perp} (\mathbf{R}(j) ) = & A _0 \cos \theta_j \mathbf{n_{\perp}} 
e^{i \mathbf{q} \cdot \mathbf{R}(j)} \nonumber \\ 
= & \frac{A_0}{2} \mathbf{n_{\perp}}  ( e^{i \theta_j} + e^{-i \theta_j} )  
\left( 1 + i q \frac{d_t}{2} \sin \theta_j \right).
\end{align}
where we take the direction of $\mathbf{n}$ as
$\mathbf{n_{||}} = (0,0,1)$ and $\mathbf{n_\perp} = (1,0,0)$. We also
take into account the fact that $qd_t$ is sufficiently small compared
with the unity, which means that in both cases the dominant
contribution to matrix element comes from the first term, whereas the
second term including $q = |\mathbf{q}|$ can be neglected, which is
known as the dipole approximation.  Hereafter we will consider only
the dominant terms.  The optical matrix element in tight-binding
approximation of Eq.~(\ref{wf_tot}) has the following form:
\begin{align}
\label{dipole_mx}
& \langle s_2, \mu_2, k_2 | \mathbf{A_q} \cdot \nabla|s_1, \mu_1, k_1 \rangle \nonumber \\
&=  \frac{1}{N} \sum_{\ell, \ell'=A,B} C^{s_2*}_{k_2\mu_2 \ell'} C^{s_1}_{k_1\mu_1\ell} 
\sum_{j,j'}  e^{-i \mathbf{k}_2 \cdot \mathbf{R}(j')} e^{i \mathbf{k}_1 \mathbf{R}(j)} \nonumber \\
& \quad\times \langle j',\ell'|\mathbf{A_q}(\mathbf{R}(j)) \cdot \nabla|j,\ell \rangle,
\end{align}
where
$| j,\ell \rangle = \chi (\mathbf{R}(j) - \mathbf{r}_\ell -
\mathbf{r})$ is the bra-ket form for the atomic orbital introduced in
Eq.~(\ref{atom_orbit}). Let us discuss Eq.~(\ref{dipole_mx}) for the
two cases of parallel and perpendicular polarization one by one.

\subsection{Perpendicular polarization}

When we put Eq.~(\ref{vect_perp}) to Eq.~(\ref{dipole_mx}), we get:
\begin{align}
\label{matrix_perp_long}
& \langle s_2, \mu_2, k_2 | \mathbf{A_q} \cdot \nabla | s_1, \mu_1, k_1 \rangle \nonumber  \\ 
= & \frac{A_0}{N} \sum_{\ell, \ell'=A,B}
C^{s_2*}_{k_2\mu_2 \ell'}  C^{s_1}_{k_1\mu_1 \ell} \nonumber \\
&\times \sum_{j,j'}
\mathbf{n_\perp} \cdot \langle j',\ell'| \nabla| j,\ell \rangle
\nonumber  \\ 
& \times  \frac{1}{2} \left( e^{i(\mathbf{k}_1 \cdot \mathbf{R}
  (j) - \mathbf{k}_2 \cdot \mathbf{R}(j') -\theta_j )} \right. \nonumber \\
 & \left. + e^{i(\mathbf{k}_1 \cdot
  \mathbf{R} (j) - \mathbf{k}_2 \cdot \mathbf{R}(j') +\theta_j )}\right) \nonumber \\ 
= & \frac{A_0}{N} \sum_{\ell, \ell'=A,B} 
C^{s_2*}_{k_2\mu_2 \ell'}  C^{s_1}_{k_1\mu_1 \ell} \nonumber  \\  
&\times \sum_{j} \frac{1}{2} \left( e^{i((\mathbf{k}_1 - \mathbf{k}_2) \cdot \mathbf{R}(j)
  -\theta_j )} \right. \nonumber \\
  & \left.+ e^{i((\mathbf{k}_1 - \mathbf{k}_2)\cdot \mathbf{R}(j) +\theta_j)}\right)  \nonumber  \\ 
& \times  \sum_{j'} \mathbf{n_\perp} \cdot
\langle j',\ell'| \nabla| j,\ell \rangle \nonumber \\
& \times e^{-i \mathbf{k}_2 \cdot (
  \mathbf{R}(j') - \mathbf{R}(j) )}.
\end{align}
Here we define two-dimensional unit vectors originated from carbon nanotube 
lattice vectors $\mathbf{C}_h, \mathbf{T}$ \cite{Sato2017}:
\begin{align}
\label{unit_vectors}
\mathbf{e}_C & = \frac{\mathbf{C}_h}{|\mathbf{C}_h|} = \frac{\mathbf{K}_1}{|\mathbf{K}_1|},  \nonumber \\
\mathbf{e}_T & = \frac{\mathbf{T}}{|\mathbf{T}|} = \frac{\mathbf{K}_2}{|\mathbf{K}_2|}.
\end{align}
Then vectors $\mathbf{k}_1,\mathbf{k}_2$ and $ \mathbf{R}(j)$ can be expressed 
by $\mathbf{e}_C$ and $\mathbf{e}_T$ as follows:
\begin{align}
\label{vectors_not}
\mathbf{k}_1 & = \mu_1 |\mathbf{K}_1| \mathbf{e}_C + k_1 \mathbf{e}_T,  \nonumber \\
\mathbf{k}_2 & = \mu_2 |\mathbf{K}_1| \mathbf{e}_C + k_2 \mathbf{e}_T, \nonumber \\ 
\mathbf{R}(j) & = \frac{\theta_j}{ |\mathbf{K}_1|} \mathbf{e}_C + R_z(j) \mathbf{e}_T.
\end{align}
Using Eq.~(\ref{vectors_not}) we simplify the phase in Eq.~(\ref{matrix_perp_long}):
\begin{equation}
\label{phase_perp}
(\mathbf{k}_1 - \mathbf{k}_2)\cdot \mathbf{R}(j) \pm \theta_j = (k_1 - k_2) R_z(j) + (\mu_1 - \mu_2 \pm 1) \theta_j
\end{equation}
Taking the summation on $j$ in Eq.~(\ref{matrix_perp_long}) we get $\delta(k_2-k_1)$ and $\delta(\mu_2-\mu_1\pm1)$.
Finally the optical matrix elment takes the following form:
\begin{align}
\label{matrix_perp_fin}
& \langle s_1, \mu_1, k_1| \mathbf{A_q} \cdot \nabla |s_2, \mu_2, k_2 \rangle  \nonumber \\
= & A_0 \sum_{\ell, \ell'=A,B} C^{s_2*}_{k_2\mu_2 \ell'} C^{s_1}_{k_1\mu_1 \ell}  \delta(k_1 - k_2) \nonumber \\
& \times  \frac{1}{2} \left( \delta(\mu_1 - \mu_2 -1) + \delta(\mu_1 - \mu_2 +1) \right) \nonumber \\ 
& \times  \sum_{j}  \mathbf{n_\perp}  \cdot \langle j,\ell'| \nabla| 0,\ell \rangle e^{-i \mathbf{k}_2 \cdot \mathbf{R}(j)}.
\end{align}

\subsection{Parallel polarization}

Similarly for parallel polarization, when we put Eq.~(\ref{vect_par})
in Eq.~(\ref{dipole_mx}), we get:
\begin{align}
\label{matrix_par_long}
& A_0 \mathbf{n_{||}} \langle s_2, \mu_2, k_2 | \nabla|s_1, \mu_1, k_1 \rangle  \nonumber \\ 
= &\frac{A_0}{N} \sum_{\ell, \ell'=A,B} C^{s_2*}_{k_2\mu_2 \ell'} C^{s_1}_{k_1\mu_1 \ell}  \nonumber \\
&\times \sum_{\mathbf{j,j'}} e^{-i \mathbf{k}_2 \cdot \mathbf{R} (j')} e^{i \mathbf{k}_1 \cdot \mathbf{R}(j)}  \nonumber \\
& \times \mathbf{n_{||}} \cdot  \langle j',\ell'| \nabla| j,\ell \rangle   \nonumber \\ 
= & \frac{A_0}{N} \sum_{\ell, \ell'=A,B}  C^{s_2*}_{k_2\mu_2 \ell'} C^{s_1}_{k_1\mu_1 \ell} \nonumber \\
& \times \sum_{j}   e^{i((\mathbf{k}_1 - \mathbf{k}_2) \cdot \mathbf{R}(j))} \nonumber \\
&  \times \sum_{j'}  \mathbf{n_{||}}  \cdot \langle j',\ell'| \nabla| j, \ell \rangle \nonumber \\
& \times e^{-i \mathbf{k}_2 \cdot ( \mathbf{R}(\mathbf{j'}) - \mathbf{R}(j) )}.
\end{align}
Using Eqs.~(\ref{unit_vectors})--\eqref{phase_perp} we finally get:
\begin{align}
\label{matrix_par_fin}
& A_0 \mathbf{n_{||}} \langle s_2, \mu_2, k_2 | \nabla|s_1, \mu_1, k_1 \rangle  \nonumber \\ 
= & A_0 \sum_{\ell, \ell'=A,B} C^{s_2*}_{k_2\mu_2 \ell'} C^{s_1}_{k_1\mu_1 \ell}  \nonumber \\
& \times \delta(k_1 - k_2) \delta(\mu_1 - \mu_2)  \nonumber \\
& \times \sum_{j}  \mathbf{n_{||}}  \cdot \langle j,\ell'| \nabla| 0, \ell \rangle  e^{-i \mathbf{k}_2 \cdot \mathbf{R}(j)} .
\end{align}

\section{Different plasmon branches}
\label{plasm}

\begin{figure}[t!]
  \centering \includegraphics[clip,width=8cm]{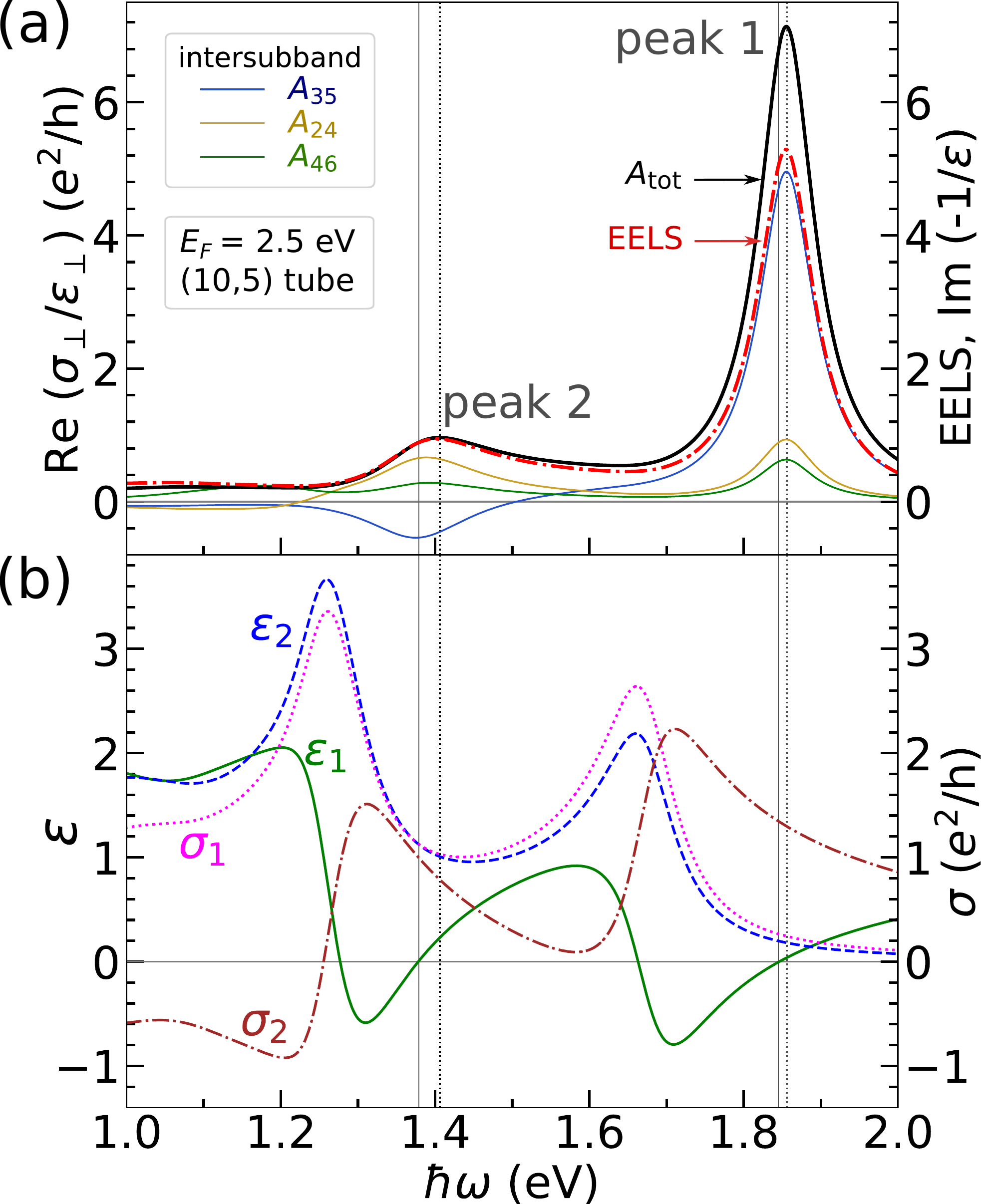}
  \caption{\label{fig:10} (a) Absorption spectra for $(10,5)$ doped
    SWNT with $E_F=2.5$ eV.  Black bold solid line represents the
    total absorption $A_{tot}$. Colored solid lines correspond to the
    dominant $A_{35}$, $A_{24}$ and $A_{46}$ intersubband
    contributions.  The EELS spectrum, $\mathrm{Im}(-1/\varepsilon)$,
    is plotted with red dash-dotted line.  (b) Real and imaginary
    parts of dielectric function and conductivity for $(10,5)$ doped
    SWNT with $E_F=1.5$ eV.  Solid vertical line corresponds to
    $\text{Re}(\varepsilon) = 0$, while dotted vertical line
    corresponds to $\rm{max}(A_\text{tot})$.}
\end{figure}

\begin{figure}[t!]
  \centering \includegraphics[clip,width=8.5cm]{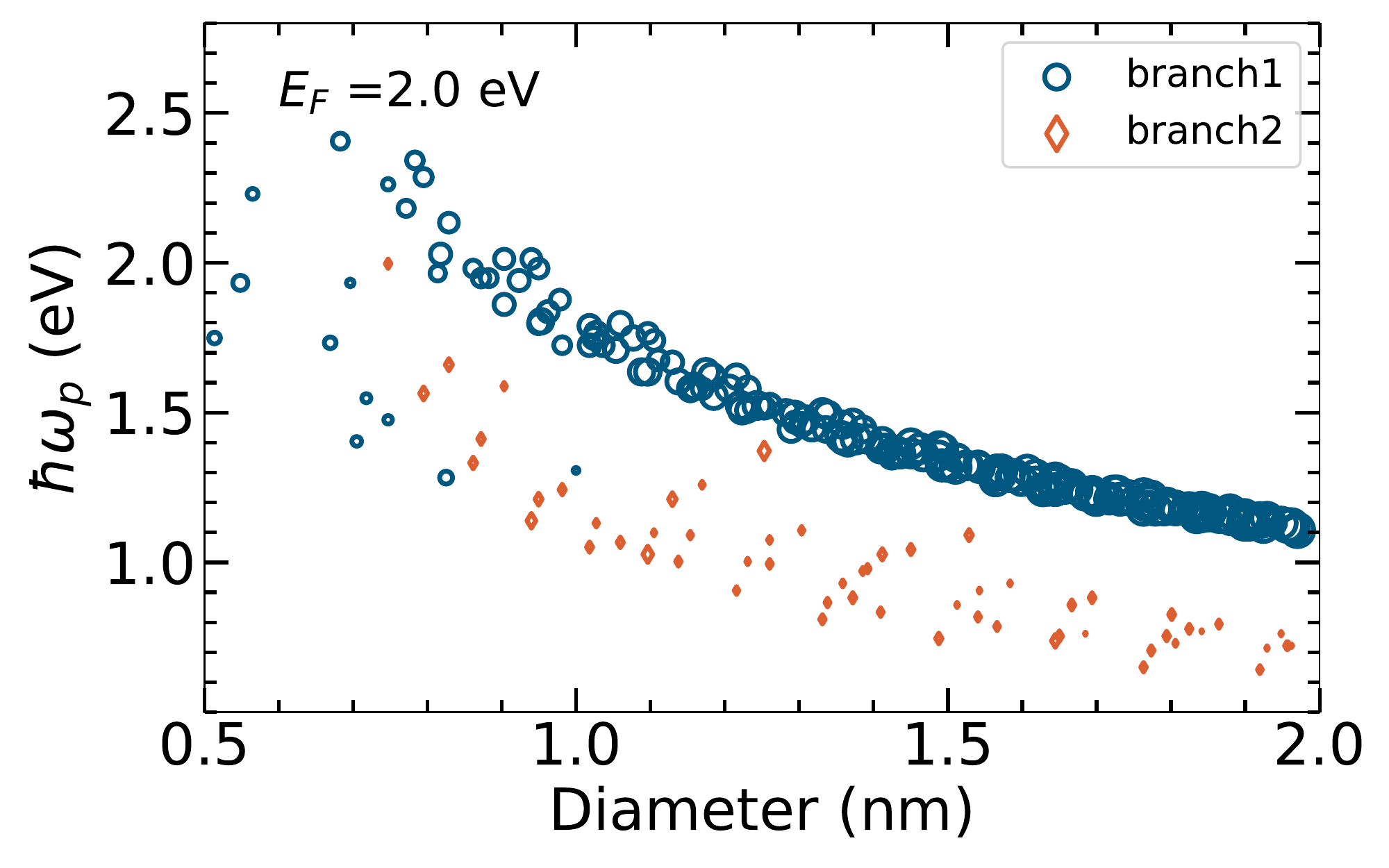}
  \caption{\label{fig:11} Two plasmon branches in doped SWNTs. Blue
    circles correspond to the main branch discussed in
    Sec.~\ref{sec:res}, orange diamonds correspond to the second
    branch plasmon, which appear at higher doping levels.  The size of
    the marker corresponds to the peak intensity.}
\end{figure}
In Sec.~\ref{sec:res} we discuss plasmon spectra only for major
plasmons, which appear first and remain dominant in terms of its
magnitude.  However, for $E_F>2.0$~eV there exist another plasmon at
the lower frequency as shown in Fig.~\ref{fig:4}~(a).  Now in
Fig.~\ref{fig:10}~(a) we plot the absorption spectra
$A_{\text{tot}}=\text{Re}(\sigma_\perp/\varepsilon_\perp)$, as well as
EELS spectra by $\text{Im}(-1/\varepsilon_\perp)$ (dash-dotted line),
as a function of photon energy for the $(10,5)$ SWNT at $E_F=2.5$
eV. We can see two prominent peaks at $1.86$ eV (peak 1) and $1.4$ eV
(peak 2), which differ by the dominant contributions
[Fig.~\ref{fig:4}~(a)], i.e., P$_{35}$ (from $A_{35}$) and P$_{24}$
(from $A_{24}$), respectively .  In particular, for the peak 2, the
absorption $A_{35}$, which is dominant for the peak 1, gives the
negative contribution.  This leads to a different behavior of the peak
2 as a function of $E_F$.

In Fig.~\ref{fig:10}~(b), we plot
$\varepsilon_1= \text{Re}(\varepsilon)$,
$\varepsilon_2=\text{Im} (\varepsilon)$,
$\sigma_1=\mathrm{Re}(\sigma)$, and $\sigma_2=\mathrm{Im}(\sigma)$ as
a function of photon energy.  The condition on plasmon excitation is
satisfied at two zero points of the real part of dielectric function
(solid vertical line). The absorption maxima (dotted vertical line)
are red-shifted regarding to $\text{Re}(\varepsilon)=0$, the shift is
larger for peak 2, since $\varepsilon_2$ is steeper around
$\omega_{p2}$.  Here we can clearly observe the effect of
$\varepsilon_2$ on plasmonic spectra:
$A_1/A_2 \propto
\varepsilon_2(\omega_{p2})/\varepsilon_2(\omega_{p1})$, where we
denote $A_1$ and $A_2$ as the intensities of plasmon peaks 1 and 2.
The presence of the second branch of intersubband plasmon have not
been mentioned any of previous works of SWNTs.  However in recent
years, several \emph{ab initio} studies show the similar second branch
for bilayer graphene, nanoribbons, and other 2D materials
\cite{Pisarra2016,Gomez2016, Torbatian2017, Torbatian2018}. The
intraband nature of the second branch plasmon in graphene nanoribbons
was supposed by Gomez~\emph{et al.}~\cite{Gomez2016}, which is
consistent with our results.  We plot both plasmon branches for SWNT
in Fig.~\ref{fig:11} for different chiralities of SWNT with $d_t<2$~nm
at $E_F=2.0$~eV.  The lower plasmon peak P$_{24}$ shows a larger
chiral angle dependence since it comes from the cutting lines pairs
closer to the $K$ point than the major plasmon P$_{35}$. Thus the
similar spreading character is observed for small-diameter SWNTs
($d_t<1$ nm) and the second branch plasmon for bigger SWNTs
($1 < d_t <2$ nm).


\begin{thebibliography}{58}%
\makeatletter
\providecommand \@ifxundefined [1]{%
 \@ifx{#1\undefined}
}%
\providecommand \@ifnum [1]{%
 \ifnum #1\expandafter \@firstoftwo
 \else \expandafter \@secondoftwo
 \fi
}%
\providecommand \@ifx [1]{%
 \ifx #1\expandafter \@firstoftwo
 \else \expandafter \@secondoftwo
 \fi
}%
\providecommand \natexlab [1]{#1}%
\providecommand \enquote  [1]{``#1''}%
\providecommand \bibnamefont  [1]{#1}%
\providecommand \bibfnamefont [1]{#1}%
\providecommand \citenamefont [1]{#1}%
\providecommand \href@noop [0]{\@secondoftwo}%
\providecommand \href [0]{\begingroup \@sanitize@url \@href}%
\providecommand \@href[1]{\@@startlink{#1}\@@href}%
\providecommand \@@href[1]{\endgroup#1\@@endlink}%
\providecommand \@sanitize@url [0]{\catcode `\\12\catcode `\$12\catcode
  `\&12\catcode `\#12\catcode `\^12\catcode `\_12\catcode `\%12\relax}%
\providecommand \@@startlink[1]{}%
\providecommand \@@endlink[0]{}%
\providecommand \url  [0]{\begingroup\@sanitize@url \@url }%
\providecommand \@url [1]{\endgroup\@href {#1}{\urlprefix }}%
\providecommand \urlprefix  [0]{URL }%
\providecommand \Eprint [0]{\href }%
\providecommand \doibase [0]{http://dx.doi.org/}%
\providecommand \selectlanguage [0]{\@gobble}%
\providecommand \bibinfo  [0]{\@secondoftwo}%
\providecommand \bibfield  [0]{\@secondoftwo}%
\providecommand \translation [1]{[#1]}%
\providecommand \BibitemOpen [0]{}%
\providecommand \bibitemStop [0]{}%
\providecommand \bibitemNoStop [0]{.\EOS\space}%
\providecommand \EOS [0]{\spacefactor3000\relax}%
\providecommand \BibitemShut  [1]{\csname bibitem#1\endcsname}%
\let\auto@bib@innerbib\@empty
\bibitem [{\citenamefont {Kataura}\ \emph {et~al.}(1999)\citenamefont
  {Kataura}, \citenamefont {Kumazawa}, \citenamefont {Maniwa}, \citenamefont
  {Umezu}, \citenamefont {Suzuki}, \citenamefont {Ohtsuka},\ and\ \citenamefont
  {Achiba}}]{Kataura1999}%
  \BibitemOpen
  \bibfield  {author} {\bibinfo {author} {\bibfnamefont {H.}~\bibnamefont
  {Kataura}}, \bibinfo {author} {\bibfnamefont {Y.}~\bibnamefont {Kumazawa}},
  \bibinfo {author} {\bibfnamefont {Y.}~\bibnamefont {Maniwa}}, \bibinfo
  {author} {\bibfnamefont {I.}~\bibnamefont {Umezu}}, \bibinfo {author}
  {\bibfnamefont {S.}~\bibnamefont {Suzuki}}, \bibinfo {author} {\bibfnamefont
  {Y.}~\bibnamefont {Ohtsuka}}, \ and\ \bibinfo {author} {\bibfnamefont
  {Y.}~\bibnamefont {Achiba}},\ }\bibfield  {title} {\enquote {\bibinfo {title}
  {Optical properties of single-wall carbon nanotubes},}\ }\href@noop {}
  {\bibfield  {journal} {\bibinfo  {journal} {Synth. Met.}\ }\textbf {\bibinfo
  {volume} {103}},\ \bibinfo {pages} {2555} (\bibinfo {year}
  {1999})}\BibitemShut {NoStop}%
\bibitem [{\citenamefont {Saito}\ \emph {et~al.}(2000)\citenamefont {Saito},
  \citenamefont {Dresselhaus},\ and\ \citenamefont {Dresselhaus}}]{Saito2000}%
  \BibitemOpen
  \bibfield  {author} {\bibinfo {author} {\bibfnamefont {R.}~\bibnamefont
  {Saito}}, \bibinfo {author} {\bibfnamefont {G.}~\bibnamefont {Dresselhaus}},
  \ and\ \bibinfo {author} {\bibfnamefont {M.~S.}\ \bibnamefont
  {Dresselhaus}},\ }\bibfield  {title} {\enquote {\bibinfo {title} {Trigonal
  warping effect of carbon nanotubes},}\ }\href@noop {} {\bibfield  {journal}
  {\bibinfo  {journal} {Phys. Rev. B}\ }\textbf {\bibinfo {volume} {61}},\
  \bibinfo {pages} {2981} (\bibinfo {year} {2000})}\BibitemShut {NoStop}%
\bibitem [{\citenamefont {Bachilo}\ \emph {et~al.}(2002)\citenamefont
  {Bachilo}, \citenamefont {Strano}, \citenamefont {Kittrell}, \citenamefont
  {Hauge}, \citenamefont {Smalley},\ and\ \citenamefont
  {Weisman}}]{Weisman2002}%
  \BibitemOpen
  \bibfield  {author} {\bibinfo {author} {\bibfnamefont {S.~M.}\ \bibnamefont
  {Bachilo}}, \bibinfo {author} {\bibfnamefont {M.~S.}\ \bibnamefont {Strano}},
  \bibinfo {author} {\bibfnamefont {C.}~\bibnamefont {Kittrell}}, \bibinfo
  {author} {\bibfnamefont {R.~H.}\ \bibnamefont {Hauge}}, \bibinfo {author}
  {\bibfnamefont {R.~E.}\ \bibnamefont {Smalley}}, \ and\ \bibinfo {author}
  {\bibfnamefont {R.~B.}\ \bibnamefont {Weisman}},\ }\bibfield  {title}
  {\enquote {\bibinfo {title} {Structure-assigned optical spectra of
  single-walled carbon nanotubes},}\ }\href@noop {} {\bibfield  {journal}
  {\bibinfo  {journal} {Science}\ }\textbf {\bibinfo {volume} {298}},\ \bibinfo
  {pages} {2361} (\bibinfo {year} {2002})}\BibitemShut {NoStop}%
\bibitem [{\citenamefont {Weisman}\ and\ \citenamefont
  {Bachilo}(2003)}]{Weisman2003}%
  \BibitemOpen
  \bibfield  {author} {\bibinfo {author} {\bibfnamefont {R.~B.}\ \bibnamefont
  {Weisman}}\ and\ \bibinfo {author} {\bibfnamefont {S.~M.}\ \bibnamefont
  {Bachilo}},\ }\bibfield  {title} {\enquote {\bibinfo {title} {Dependence of
  optical transition energies on structure for single-walled carbon nanotubes
  in aqueous suspension:  an empirical $\mathrm{Kataura}$ plot},}\
  }\href@noop {} {\bibfield  {journal} {\bibinfo  {journal} {Nano Lett.}\
  }\textbf {\bibinfo {volume} {3}},\ \bibinfo {pages} {1235} (\bibinfo {year}
  {2003})}\BibitemShut {NoStop}%
\bibitem [{\citenamefont {Avouris}\ \emph {et~al.}(2006)\citenamefont
  {Avouris}, \citenamefont {Chen}, \citenamefont {Freitag}, \citenamefont
  {Perebeinos},\ and\ \citenamefont {Tsang}}]{Avouris2006}%
  \BibitemOpen
  \bibfield  {author} {\bibinfo {author} {\bibfnamefont {Ph.}\ \bibnamefont
  {Avouris}}, \bibinfo {author} {\bibfnamefont {J.}~\bibnamefont {Chen}},
  \bibinfo {author} {\bibfnamefont {M.}~\bibnamefont {Freitag}}, \bibinfo
  {author} {\bibfnamefont {V.}~\bibnamefont {Perebeinos}}, \ and\ \bibinfo
  {author} {\bibfnamefont {J.~C.}\ \bibnamefont {Tsang}},\ }\bibfield  {title}
  {\enquote {\bibinfo {title} {Carbon nanotube optoelectronics},}\ }\href@noop
  {} {\bibfield  {journal} {\bibinfo  {journal} {Phys. Status Solidi B}\
  }\textbf {\bibinfo {volume} {243}},\ \bibinfo {pages} {3197} (\bibinfo {year}
  {2006})}\BibitemShut {NoStop}%
\bibitem [{\citenamefont {Avouris}\ \emph {et~al.}(2008)\citenamefont
  {Avouris}, \citenamefont {Freitag},\ and\ \citenamefont
  {Perebeinos}}]{Avouris2008}%
  \BibitemOpen
  \bibfield  {author} {\bibinfo {author} {\bibfnamefont {Ph.}\ \bibnamefont
  {Avouris}}, \bibinfo {author} {\bibfnamefont {M.}~\bibnamefont {Freitag}}, \
  and\ \bibinfo {author} {\bibfnamefont {V.}~\bibnamefont {Perebeinos}},\
  }\bibfield  {title} {\enquote {\bibinfo {title} {Carbon-nanotube photonics
  and optoelectronics},}\ }\href@noop {} {\bibfield  {journal} {\bibinfo
  {journal} {Nat. Photonics}\ }\textbf {\bibinfo {volume} {2}},\ \bibinfo
  {pages} {341} (\bibinfo {year} {2008})}\BibitemShut {NoStop}%
\bibitem [{\citenamefont {Kaskela}\ \emph {et~al.}(2010)\citenamefont
  {Kaskela}, \citenamefont {Nasibulin}, \citenamefont {Timmermans},
  \citenamefont {Aitchison}, \citenamefont {Papadimitratos}, \citenamefont
  {Tian}, \citenamefont {Zhu}, \citenamefont {Jiang}, \citenamefont {Brown},
  \citenamefont {Zakhidov},\ and\ \citenamefont {Kauppinen}}]{Kaskela2010}%
  \BibitemOpen
  \bibfield  {author} {\bibinfo {author} {\bibfnamefont {A.}~\bibnamefont
  {Kaskela}}, \bibinfo {author} {\bibfnamefont {A.~G.}\ \bibnamefont
  {Nasibulin}}, \bibinfo {author} {\bibfnamefont {M.~Y.}\ \bibnamefont
  {Timmermans}}, \bibinfo {author} {\bibfnamefont {B.}~\bibnamefont
  {Aitchison}}, \bibinfo {author} {\bibfnamefont {A.}~\bibnamefont
  {Papadimitratos}}, \bibinfo {author} {\bibfnamefont {Y.}~\bibnamefont
  {Tian}}, \bibinfo {author} {\bibfnamefont {Z.}~\bibnamefont {Zhu}}, \bibinfo
  {author} {\bibfnamefont {H.}~\bibnamefont {Jiang}}, \bibinfo {author}
  {\bibfnamefont {D.~P.}\ \bibnamefont {Brown}}, \bibinfo {author}
  {\bibfnamefont {A.}~\bibnamefont {Zakhidov}}, \ and\ \bibinfo {author}
  {\bibfnamefont {E.~I.}\ \bibnamefont {Kauppinen}},\ }\bibfield  {title}
  {\enquote {\bibinfo {title} {Aerosol-synthesized swcnt networks with tunable
  conductivity and transparency by a dry transfer technique},}\ }\href
  {\doibase 10.1021/nl101680s} {\bibfield  {journal} {\bibinfo  {journal} {Nano
  Lett.}\ }\textbf {\bibinfo {volume} {10}},\ \bibinfo {pages} {4349} (\bibinfo
  {year} {2010})}\BibitemShut {NoStop}%
\bibitem [{\citenamefont {Tsapenko}\ \emph {et~al.}(2018)\citenamefont
  {Tsapenko}, \citenamefont {Goldt}, \citenamefont {Shulga}, \citenamefont
  {Popov}, \citenamefont {Maslakov}, \citenamefont {Anisimov}, \citenamefont
  {Sorokin},\ and\ \citenamefont {Nasibulin}}]{Tsapenko2018}%
  \BibitemOpen
  \bibfield  {author} {\bibinfo {author} {\bibfnamefont {A.~P.}\ \bibnamefont
  {Tsapenko}}, \bibinfo {author} {\bibfnamefont {A.~E.}\ \bibnamefont {Goldt}},
  \bibinfo {author} {\bibfnamefont {E.}~\bibnamefont {Shulga}}, \bibinfo
  {author} {\bibfnamefont {Z.~I.}\ \bibnamefont {Popov}}, \bibinfo {author}
  {\bibfnamefont {K.~I.}\ \bibnamefont {Maslakov}}, \bibinfo {author}
  {\bibfnamefont {A.~S.}\ \bibnamefont {Anisimov}}, \bibinfo {author}
  {\bibfnamefont {P.~B.}\ \bibnamefont {Sorokin}}, \ and\ \bibinfo {author}
  {\bibfnamefont {A.~G.}\ \bibnamefont {Nasibulin}},\ }\bibfield  {title}
  {\enquote {\bibinfo {title} {Highly conductive and transparent films of
  $\mathrm{HAuCl}_4$-doped single-walled carbon nanotubes for flexible
  applications},}\ }\href {\doibase 10.1016/j.carbon.2018.01.016} {\bibfield
  {journal} {\bibinfo  {journal} {Carbon}\ }\textbf {\bibinfo {volume} {130}},\
  \bibinfo {pages} {448} (\bibinfo {year} {2018})}\BibitemShut {NoStop}%
\bibitem [{\citenamefont {Ajiki}\ and\ \citenamefont {Ando}(1994)}]{Ajiki94}%
  \BibitemOpen
  \bibfield  {author} {\bibinfo {author} {\bibfnamefont {H.}~\bibnamefont
  {Ajiki}}\ and\ \bibinfo {author} {\bibfnamefont {T.}~\bibnamefont {Ando}},\
  }\bibfield  {title} {\enquote {\bibinfo {title}
  {$\mathrm{Aharonov}$-$\mathrm{Bohm}$ effect in carbon nanotubes},}\ }\href
  {\doibase 10.1016/0921-4526(94)91112-6} {\bibfield  {journal} {\bibinfo
  {journal} {Phys. B}\ }\textbf {\bibinfo {volume} {201}},\ \bibinfo {pages}
  {349} (\bibinfo {year} {1994})}\BibitemShut {NoStop}%
\bibitem [{\citenamefont {Hwang}\ \emph {et~al.}(2000)\citenamefont {Hwang},
  \citenamefont {Gommans}, \citenamefont {Ugawa}, \citenamefont {Tashiro},
  \citenamefont {Haggenmueller}, \citenamefont {Winey}, \citenamefont
  {Fischer}, \citenamefont {Tanner},\ and\ \citenamefont
  {Rinzler}}]{Hwang2000}%
  \BibitemOpen
  \bibfield  {author} {\bibinfo {author} {\bibfnamefont {J.}~\bibnamefont
  {Hwang}}, \bibinfo {author} {\bibfnamefont {H.~H.}\ \bibnamefont {Gommans}},
  \bibinfo {author} {\bibfnamefont {A.}~\bibnamefont {Ugawa}}, \bibinfo
  {author} {\bibfnamefont {H.}~\bibnamefont {Tashiro}}, \bibinfo {author}
  {\bibfnamefont {R.}~\bibnamefont {Haggenmueller}}, \bibinfo {author}
  {\bibfnamefont {K.~I.}\ \bibnamefont {Winey}}, \bibinfo {author}
  {\bibfnamefont {J.~E.}\ \bibnamefont {Fischer}}, \bibinfo {author}
  {\bibfnamefont {D.~B.}\ \bibnamefont {Tanner}}, \ and\ \bibinfo {author}
  {\bibfnamefont {A.~G.}\ \bibnamefont {Rinzler}},\ }\bibfield  {title}
  {\enquote {\bibinfo {title} {Polarized spectroscopy of aligned single-wall
  carbon nanotubes},}\ }\href@noop {} {\bibfield  {journal} {\bibinfo
  {journal} {Phys. Rev. B}\ }\textbf {\bibinfo {volume} {62}},\ \bibinfo
  {pages} {R13310} (\bibinfo {year} {2000})}\BibitemShut {NoStop}%
\bibitem [{\citenamefont {Jiang}\ \emph {et~al.}(2004)\citenamefont {Jiang},
  \citenamefont {Saito}, \citenamefont {Gr\"uneis}, \citenamefont
  {Dresselhaus},\ and\ \citenamefont {Dresselhaus}}]{Jiang2004}%
  \BibitemOpen
  \bibfield  {author} {\bibinfo {author} {\bibfnamefont {J.}~\bibnamefont
  {Jiang}}, \bibinfo {author} {\bibfnamefont {R.}~\bibnamefont {Saito}},
  \bibinfo {author} {\bibfnamefont {A.}~\bibnamefont {Gr\"uneis}}, \bibinfo
  {author} {\bibfnamefont {G.}~\bibnamefont {Dresselhaus}}, \ and\ \bibinfo
  {author} {\bibfnamefont {M.S.}\ \bibnamefont {Dresselhaus}},\ }\bibfield
  {title} {\enquote {\bibinfo {title} {Optical absorption matrix elements in
  single-wall carbon nanotubes},}\ }\href {\doibase
  10.1016/j.carbon.2004.07.028} {\bibfield  {journal} {\bibinfo  {journal}
  {Carbon}\ }\textbf {\bibinfo {volume} {42}},\ \bibinfo {pages} {3169}
  (\bibinfo {year} {2004})}\BibitemShut {NoStop}%
\bibitem [{\citenamefont {Murakami}\ \emph {et~al.}(2005)\citenamefont
  {Murakami}, \citenamefont {Einarsson}, \citenamefont {Edamura},\ and\
  \citenamefont {Maruyama}}]{Murakami2005}%
  \BibitemOpen
  \bibfield  {author} {\bibinfo {author} {\bibfnamefont {Y.}~\bibnamefont
  {Murakami}}, \bibinfo {author} {\bibfnamefont {E.}~\bibnamefont {Einarsson}},
  \bibinfo {author} {\bibfnamefont {T.}~\bibnamefont {Edamura}}, \ and\
  \bibinfo {author} {\bibfnamefont {S.}~\bibnamefont {Maruyama}},\ }\bibfield
  {title} {\enquote {\bibinfo {title} {Polarization dependence of the optical
  absorption of single-walled carbon nanotubes},}\ }\href@noop {} {\bibfield
  {journal} {\bibinfo  {journal} {Phys. Rev. Lett.}\ }\textbf {\bibinfo
  {volume} {94}},\ \bibinfo {pages} {087402} (\bibinfo {year}
  {2005})}\BibitemShut {NoStop}%
\bibitem [{\citenamefont {Li}\ \emph {et~al.}(2001)\citenamefont {Li},
  \citenamefont {Tang}, \citenamefont {Liu}, \citenamefont {Wang},
  \citenamefont {Chan}, \citenamefont {Saito}, \citenamefont {Okada},
  \citenamefont {Li}, \citenamefont {Chen}, \citenamefont {Nagasawa},\ and\
  \citenamefont {Tsuda}}]{Li2001}%
  \BibitemOpen
  \bibfield  {author} {\bibinfo {author} {\bibfnamefont {Z.~M.}\ \bibnamefont
  {Li}}, \bibinfo {author} {\bibfnamefont {Z.~K.}\ \bibnamefont {Tang}},
  \bibinfo {author} {\bibfnamefont {H.~J.}\ \bibnamefont {Liu}}, \bibinfo
  {author} {\bibfnamefont {N.}~\bibnamefont {Wang}}, \bibinfo {author}
  {\bibfnamefont {C.~T.}\ \bibnamefont {Chan}}, \bibinfo {author}
  {\bibfnamefont {R.}~\bibnamefont {Saito}}, \bibinfo {author} {\bibfnamefont
  {S.}~\bibnamefont {Okada}}, \bibinfo {author} {\bibfnamefont {G.~D.}\
  \bibnamefont {Li}}, \bibinfo {author} {\bibfnamefont {J.~S.}\ \bibnamefont
  {Chen}}, \bibinfo {author} {\bibfnamefont {N.}~\bibnamefont {Nagasawa}}, \
  and\ \bibinfo {author} {\bibfnamefont {S.}~\bibnamefont {Tsuda}},\ }\bibfield
   {title} {\enquote {\bibinfo {title} {Polarized absorption spectra of
  single-walled 4 $\mathrm{{\AA}}$ carbon nanotubes aligned in channels of an
  $\mathrm{AlPO}_4$--$5$ single crystal},}\ }\href {\doibase
  10.1103/PhysRevLett.87.127401} {\bibfield  {journal} {\bibinfo  {journal}
  {Phys. Rev. Lett.}\ }\textbf {\bibinfo {volume} {87}},\ \bibinfo {pages}
  {127401} (\bibinfo {year} {2001})}\BibitemShut {NoStop}%
\bibitem [{\citenamefont {Uryu}\ and\ \citenamefont {Ando}(2006)}]{Uryu2006}%
  \BibitemOpen
  \bibfield  {author} {\bibinfo {author} {\bibfnamefont {S.}~\bibnamefont
  {Uryu}}\ and\ \bibinfo {author} {\bibfnamefont {T.}~\bibnamefont {Ando}},\
  }\bibfield  {title} {\enquote {\bibinfo {title} {Exciton absorption of
  perpendicularly polarized light in carbon nanotubes},}\ }\href@noop {}
  {\bibfield  {journal} {\bibinfo  {journal} {Phys. Rev. B}\ }\textbf {\bibinfo
  {volume} {74}},\ \bibinfo {pages} {155411} (\bibinfo {year}
  {2006})}\BibitemShut {NoStop}%
\bibitem [{\citenamefont {Spataru}\ \emph {et~al.}(2004)\citenamefont
  {Spataru}, \citenamefont {Ismail-Beigi}, \citenamefont {Benedict},\ and\
  \citenamefont {Louie}}]{Spataru2004}%
  \BibitemOpen
  \bibfield  {author} {\bibinfo {author} {\bibfnamefont {C.~D.}\ \bibnamefont
  {Spataru}}, \bibinfo {author} {\bibfnamefont {S.}~\bibnamefont
  {Ismail-Beigi}}, \bibinfo {author} {\bibfnamefont {L.~X.}\ \bibnamefont
  {Benedict}}, \ and\ \bibinfo {author} {\bibfnamefont {S.~G.}\ \bibnamefont
  {Louie}},\ }\bibfield  {title} {\enquote {\bibinfo {title} {Excitonic effects
  and optical spectra of single-walled carbon nanotubes},}\ }\href@noop {}
  {\bibfield  {journal} {\bibinfo  {journal} {Phys. Rev. Lett.}\ }\textbf
  {\bibinfo {volume} {92}},\ \bibinfo {pages} {077402} (\bibinfo {year}
  {2004})}\BibitemShut {NoStop}%
\bibitem [{\citenamefont {Wang}(2005)}]{Wang2005}%
  \BibitemOpen
  \bibfield  {author} {\bibinfo {author} {\bibfnamefont {F.}~\bibnamefont
  {Wang}},\ }\bibfield  {title} {\enquote {\bibinfo {title} {The optical
  resonances in carbon nanotubes arise from excitons},}\ }\href {\doibase
  10.1126/science.1110265} {\bibfield  {journal} {\bibinfo  {journal}
  {Science}\ }\textbf {\bibinfo {volume} {308}},\ \bibinfo {pages} {838}
  (\bibinfo {year} {2005})}\BibitemShut {NoStop}%
\bibitem [{\citenamefont {Dukovic}\ \emph {et~al.}(2005)\citenamefont
  {Dukovic}, \citenamefont {Wang}, \citenamefont {Song}, \citenamefont {Sfeir},
  \citenamefont {Heinz},\ and\ \citenamefont {Brus}}]{Dukovic2005}%
  \BibitemOpen
  \bibfield  {author} {\bibinfo {author} {\bibfnamefont {G.}~\bibnamefont
  {Dukovic}}, \bibinfo {author} {\bibfnamefont {F.}~\bibnamefont {Wang}},
  \bibinfo {author} {\bibfnamefont {D.}~\bibnamefont {Song}}, \bibinfo {author}
  {\bibfnamefont {M.~Y.}\ \bibnamefont {Sfeir}}, \bibinfo {author}
  {\bibfnamefont {T.~F.}\ \bibnamefont {Heinz}}, \ and\ \bibinfo {author}
  {\bibfnamefont {L.~E.}\ \bibnamefont {Brus}},\ }\bibfield  {title} {\enquote
  {\bibinfo {title} {Structural dependence of excitonic optical transitions and
  band-gap energies in carbon nanotubes},}\ }\href@noop {} {\bibfield
  {journal} {\bibinfo  {journal} {Nano Lett.}\ }\textbf {\bibinfo {volume}
  {5}},\ \bibinfo {pages} {2314} (\bibinfo {year} {2005})}\BibitemShut
  {NoStop}%
\bibitem [{\citenamefont {Jiang}\ \emph {et~al.}(2007)\citenamefont {Jiang},
  \citenamefont {Saito}, \citenamefont {Sato}, \citenamefont {Park},
  \citenamefont {Samsonidze}, \citenamefont {Jorio}, \citenamefont
  {Dresselhaus},\ and\ \citenamefont {Dresselhaus}}]{Jiang2007}%
  \BibitemOpen
  \bibfield  {author} {\bibinfo {author} {\bibfnamefont {J.}~\bibnamefont
  {Jiang}}, \bibinfo {author} {\bibfnamefont {R.}~\bibnamefont {Saito}},
  \bibinfo {author} {\bibfnamefont {K.}~\bibnamefont {Sato}}, \bibinfo {author}
  {\bibfnamefont {J.~S.}\ \bibnamefont {Park}}, \bibinfo {author}
  {\bibfnamefont {Ge.~G.}\ \bibnamefont {Samsonidze}}, \bibinfo {author}
  {\bibfnamefont {A.}~\bibnamefont {Jorio}}, \bibinfo {author} {\bibfnamefont
  {G.}~\bibnamefont {Dresselhaus}}, \ and\ \bibinfo {author} {\bibfnamefont
  {M.~S.}\ \bibnamefont {Dresselhaus}},\ }\bibfield  {title} {\enquote
  {\bibinfo {title} {Exciton-photon, exciton-phonon matrix elements, and
  resonant raman intensity of single-wall carbon nanotubes},}\ }\href@noop {}
  {\bibfield  {journal} {\bibinfo  {journal} {Phys. Rev. B}\ }\textbf {\bibinfo
  {volume} {75}},\ \bibinfo {pages} {035405} (\bibinfo {year}
  {2007})}\BibitemShut {NoStop}%
\bibitem [{\citenamefont {Sasaki}\ \emph {et~al.}(2016)\citenamefont {Sasaki},
  \citenamefont {Murakami},\ and\ \citenamefont
  {Yamamoto}}]{sasaki16-interplasmon}%
  \BibitemOpen
  \bibfield  {author} {\bibinfo {author} {\bibfnamefont {K.}~\bibnamefont
  {Sasaki}}, \bibinfo {author} {\bibfnamefont {S.}~\bibnamefont {Murakami}}, \
  and\ \bibinfo {author} {\bibfnamefont {H.}~\bibnamefont {Yamamoto}},\
  }\bibfield  {title} {\enquote {\bibinfo {title} {Theory of intraband plasmons
  in doped carbon nanotubes: Rolled surface-plasmons of graphene},}\
  }\href@noop {} {\bibfield  {journal} {\bibinfo  {journal} {Appl. Phys.
  Lett.}\ }\textbf {\bibinfo {volume} {108}},\ \bibinfo {pages} {163109}
  (\bibinfo {year} {2016})}\BibitemShut {NoStop}%
\bibitem [{\citenamefont {Sasaki}\ and\ \citenamefont
  {Tokura}(2018)}]{sasaki18-plasmonpol}%
  \BibitemOpen
  \bibfield  {author} {\bibinfo {author} {\bibfnamefont {K.}~\bibnamefont
  {Sasaki}}\ and\ \bibinfo {author} {\bibfnamefont {Y.}~\bibnamefont
  {Tokura}},\ }\bibfield  {title} {\enquote {\bibinfo {title} {Theory of a
  carbon-nanotube polarization switch},}\ }\href@noop {} {\bibfield  {journal}
  {\bibinfo  {journal} {Phys. Rev. Applied}\ }\textbf {\bibinfo {volume} {9}},\
  \bibinfo {pages} {034018} (\bibinfo {year} {2018})}\BibitemShut {NoStop}%
\bibitem [{\citenamefont {Yanagi}\ \emph {et~al.}(2018)\citenamefont {Yanagi},
  \citenamefont {Okada}, \citenamefont {Ichinose}, \citenamefont {Yomogida},
  \citenamefont {Katsutani}, \citenamefont {Gao},\ and\ \citenamefont
  {Kono}}]{yanagi18-isbp}%
  \BibitemOpen
  \bibfield  {author} {\bibinfo {author} {\bibfnamefont {K.}~\bibnamefont
  {Yanagi}}, \bibinfo {author} {\bibfnamefont {R.}~\bibnamefont {Okada}},
  \bibinfo {author} {\bibfnamefont {Y.}~\bibnamefont {Ichinose}}, \bibinfo
  {author} {\bibfnamefont {Y.}~\bibnamefont {Yomogida}}, \bibinfo {author}
  {\bibfnamefont {F.}~\bibnamefont {Katsutani}}, \bibinfo {author}
  {\bibfnamefont {W.}~\bibnamefont {Gao}}, \ and\ \bibinfo {author}
  {\bibfnamefont {J.}~\bibnamefont {Kono}},\ }\bibfield  {title} {\enquote
  {\bibinfo {title} {Intersubband plasmons in the quantum limit in gated and
  aligned carbon nanotubes},}\ }\href@noop {} {\bibfield  {journal} {\bibinfo
  {journal} {Nat. Commun.}\ }\textbf {\bibinfo {volume} {9}},\ \bibinfo {pages}
  {1121} (\bibinfo {year} {2018})}\BibitemShut {NoStop}%
\bibitem [{\citenamefont {Senga}\ \emph {et~al.}(2016)\citenamefont {Senga},
  \citenamefont {Pichler},\ and\ \citenamefont {Suenaga}}]{Senga2016}%
  \BibitemOpen
  \bibfield  {author} {\bibinfo {author} {\bibfnamefont {R.}~\bibnamefont
  {Senga}}, \bibinfo {author} {\bibfnamefont {T.}~\bibnamefont {Pichler}}, \
  and\ \bibinfo {author} {\bibfnamefont {K.}~\bibnamefont {Suenaga}},\
  }\bibfield  {title} {\enquote {\bibinfo {title} {Electron spectroscopy of
  single quantum objects to directly correlate the local structure to their
  electronic transport and optical properties},}\ }\href@noop {} {\bibfield
  {journal} {\bibinfo  {journal} {Nano Lett.}\ }\textbf {\bibinfo {volume}
  {16}},\ \bibinfo {pages} {3661} (\bibinfo {year} {2016})}\BibitemShut
  {NoStop}%
\bibitem [{\citenamefont {Senga}\ \emph {et~al.}(2018)\citenamefont {Senga},
  \citenamefont {Pichler}, \citenamefont {Yomogida}, \citenamefont {Tanaka},
  \citenamefont {Kataura},\ and\ \citenamefont {Suenaga}}]{senga18-eels}%
  \BibitemOpen
  \bibfield  {author} {\bibinfo {author} {\bibfnamefont {R.}~\bibnamefont
  {Senga}}, \bibinfo {author} {\bibfnamefont {T.}~\bibnamefont {Pichler}},
  \bibinfo {author} {\bibfnamefont {Y.}~\bibnamefont {Yomogida}}, \bibinfo
  {author} {\bibfnamefont {T.}~\bibnamefont {Tanaka}}, \bibinfo {author}
  {\bibfnamefont {H.}~\bibnamefont {Kataura}}, \ and\ \bibinfo {author}
  {\bibfnamefont {K.}~\bibnamefont {Suenaga}},\ }\bibfield  {title} {\enquote
  {\bibinfo {title} {Direct proof of a defect-modulated gap transition in
  semiconducting nanotubes},}\ }\href@noop {} {\bibfield  {journal} {\bibinfo
  {journal} {Nano Lett.}\ }\textbf {\bibinfo {volume} {18}},\ \bibinfo {pages}
  {3920} (\bibinfo {year} {2018})}\BibitemShut {NoStop}%
\bibitem [{\citenamefont {Kuzuo}\ \emph {et~al.}(1992)\citenamefont {Kuzuo},
  \citenamefont {Terauchi},\ and\ \citenamefont {Tanaka}}]{Kuzuo1992}%
  \BibitemOpen
  \bibfield  {author} {\bibinfo {author} {\bibfnamefont {R.}~\bibnamefont
  {Kuzuo}}, \bibinfo {author} {\bibfnamefont {M.}~\bibnamefont {Terauchi}}, \
  and\ \bibinfo {author} {\bibfnamefont {M.}~\bibnamefont {Tanaka}},\
  }\bibfield  {title} {\enquote {\bibinfo {title} {Electron energy-loss spectra
  of carbon nanotubes},}\ }\href@noop {} {\bibfield  {journal} {\bibinfo
  {journal} {Jpn. J. Appl. Phys.}\ }\textbf {\bibinfo {volume} {31}},\ \bibinfo
  {pages} {L1484} (\bibinfo {year} {1992})}\BibitemShut {NoStop}%
\bibitem [{\citenamefont {Kuzuo}\ \emph {et~al.}(1994)\citenamefont {Kuzuo},
  \citenamefont {Terauchi}, \citenamefont {Tanaka},\ and\ \citenamefont
  {Saito}}]{Kuzuo1994}%
  \BibitemOpen
  \bibfield  {author} {\bibinfo {author} {\bibfnamefont {R.}~\bibnamefont
  {Kuzuo}}, \bibinfo {author} {\bibfnamefont {M.}~\bibnamefont {Terauchi}},
  \bibinfo {author} {\bibfnamefont {M.}~\bibnamefont {Tanaka}}, \ and\ \bibinfo
  {author} {\bibfnamefont {Y.}~\bibnamefont {Saito}},\ }\bibfield  {title}
  {\enquote {\bibinfo {title} {Electron energy-loss spectra of single-shell
  carbon nanotubes},}\ }\href@noop {} {\bibfield  {journal} {\bibinfo
  {journal} {Jpn. J. Appl. Phys.}\ }\textbf {\bibinfo {volume} {33}},\ \bibinfo
  {pages} {L1316} (\bibinfo {year} {1994})}\BibitemShut {NoStop}%
\bibitem [{\citenamefont {Papagno}\ and\ \citenamefont
  {Caputi}(1983)}]{Papagno1983}%
  \BibitemOpen
  \bibfield  {author} {\bibinfo {author} {\bibfnamefont {L.}~\bibnamefont
  {Papagno}}\ and\ \bibinfo {author} {\bibfnamefont {L.~S.}\ \bibnamefont
  {Caputi}},\ }\bibfield  {title} {\enquote {\bibinfo {title} {Electronic
  structure of graphite: Single particle and collective excitations studied by
  $\mathrm{EELS}$, $\mathrm{SEE}$ and $\mathrm{K}$ edge loss techniques},}\
  }\href {\doibase 10.1016/0039-6028(83)90583-6} {\bibfield  {journal}
  {\bibinfo  {journal} {Surf. Sci.}\ }\textbf {\bibinfo {volume} {125}},\
  \bibinfo {pages} {530} (\bibinfo {year} {1983})}\BibitemShut {NoStop}%
\bibitem [{\citenamefont {Liou}\ \emph {et~al.}(2014)\citenamefont {Liou},
  \citenamefont {Breitwieser}, \citenamefont {Chen}, \citenamefont {Pai},
  \citenamefont {Guo},\ and\ \citenamefont {Chu}}]{Liou2014}%
  \BibitemOpen
  \bibfield  {author} {\bibinfo {author} {\bibfnamefont {S.~C.}\ \bibnamefont
  {Liou}}, \bibinfo {author} {\bibfnamefont {R.}~\bibnamefont {Breitwieser}},
  \bibinfo {author} {\bibfnamefont {C.~H.}\ \bibnamefont {Chen}}, \bibinfo
  {author} {\bibfnamefont {W.~W.}\ \bibnamefont {Pai}}, \bibinfo {author}
  {\bibfnamefont {G.~Y.}\ \bibnamefont {Guo}}, \ and\ \bibinfo {author}
  {\bibfnamefont {M.~W.}\ \bibnamefont {Chu}},\ }\bibfield  {title} {\enquote
  {\bibinfo {title} {$\pi$-plasmon dispersion in free-standing monolayer
  graphene investigated by momentum-resolved electron energy-loss
  spectroscopy},}\ }\href {\doibase 10.1017/s1431927614010678} {\bibfield
  {journal} {\bibinfo  {journal} {Microsc. Microanal.}\ }\textbf {\bibinfo
  {volume} {20}},\ \bibinfo {pages} {1788} (\bibinfo {year}
  {2014})}\BibitemShut {NoStop}%
\bibitem [{\citenamefont {Hu}\ \emph {et~al.}(2014)\citenamefont {Hu},
  \citenamefont {Zeng}, \citenamefont {Wang}, \citenamefont {Li}, \citenamefont
  {Kan},\ and\ \citenamefont {Liu}}]{Hu2014}%
  \BibitemOpen
  \bibfield  {author} {\bibinfo {author} {\bibfnamefont {J.}~\bibnamefont
  {Hu}}, \bibinfo {author} {\bibfnamefont {H.}~\bibnamefont {Zeng}}, \bibinfo
  {author} {\bibfnamefont {C.}~\bibnamefont {Wang}}, \bibinfo {author}
  {\bibfnamefont {Z.}~\bibnamefont {Li}}, \bibinfo {author} {\bibfnamefont
  {C.}~\bibnamefont {Kan}}, \ and\ \bibinfo {author} {\bibfnamefont
  {Y.}~\bibnamefont {Liu}},\ }\bibfield  {title} {\enquote {\bibinfo {title}
  {Interband $\pi$ plasmon of graphene: strong small-size and field-enhancement
  effects},}\ }\href {\doibase 10.1039/c4cp02299h} {\bibfield  {journal}
  {\bibinfo  {journal} {Phys. Chem. Chem. Phys.}\ }\textbf {\bibinfo {volume}
  {16}},\ \bibinfo {pages} {23483} (\bibinfo {year} {2014})}\BibitemShut
  {NoStop}%
\bibitem [{\citenamefont {Lin}\ and\ \citenamefont {Shung}(1994)}]{Lin1994}%
  \BibitemOpen
  \bibfield  {author} {\bibinfo {author} {\bibfnamefont {M.~F.}\ \bibnamefont
  {Lin}}\ and\ \bibinfo {author} {\bibfnamefont {K.~W.-K.}\ \bibnamefont
  {Shung}},\ }\bibfield  {title} {\enquote {\bibinfo {title} {Plasmons and
  optical properties of carbon nanotubes},}\ }\href@noop {} {\bibfield
  {journal} {\bibinfo  {journal} {Phys. Rev. B}\ }\textbf {\bibinfo {volume}
  {50}},\ \bibinfo {pages} {17744} (\bibinfo {year} {1994})}\BibitemShut
  {NoStop}%
\bibitem [{\citenamefont {Lin}\ \emph {et~al.}(1996)\citenamefont {Lin},
  \citenamefont {Chuu}, \citenamefont {Huang}, \citenamefont {Lin},\ and\
  \citenamefont {Shung}}]{Lin1996}%
  \BibitemOpen
  \bibfield  {author} {\bibinfo {author} {\bibfnamefont {M.~F.}\ \bibnamefont
  {Lin}}, \bibinfo {author} {\bibfnamefont {D.~S.}\ \bibnamefont {Chuu}},
  \bibinfo {author} {\bibfnamefont {C.~S.}\ \bibnamefont {Huang}}, \bibinfo
  {author} {\bibfnamefont {Y.~K.}\ \bibnamefont {Lin}}, \ and\ \bibinfo
  {author} {\bibfnamefont {K.~W.-K.}\ \bibnamefont {Shung}},\ }\bibfield
  {title} {\enquote {\bibinfo {title} {Collective excitations in a single-layer
  carbon nanotube},}\ }\href {\doibase 10.1103/physrevb.53.15493} {\bibfield
  {journal} {\bibinfo  {journal} {Phys. Rev. B}\ }\textbf {\bibinfo {volume}
  {53}},\ \bibinfo {pages} {15493} (\bibinfo {year} {1996})}\BibitemShut
  {NoStop}%
\bibitem [{\citenamefont {de~Abajo}(2014)}]{Abajo2014}%
  \BibitemOpen
  \bibfield  {author} {\bibinfo {author} {\bibfnamefont {F.~J.~Garc{\'{\i}}a}\
  \bibnamefont {de~Abajo}},\ }\bibfield  {title} {\enquote {\bibinfo {title}
  {Graphene plasmonics: Challenges and opportunities},}\ }\href@noop {}
  {\bibfield  {journal} {\bibinfo  {journal} {ACS Photonics}\ }\textbf
  {\bibinfo {volume} {1}},\ \bibinfo {pages} {135} (\bibinfo {year}
  {2014})}\BibitemShut {NoStop}%
\bibitem [{\citenamefont {Bondarev}(2012)}]{Bondarev2012}%
  \BibitemOpen
  \bibfield  {author} {\bibinfo {author} {\bibfnamefont {I.~V.}\ \bibnamefont
  {Bondarev}},\ }\bibfield  {title} {\enquote {\bibinfo {title} {Single-wall
  carbon nanotubes as coherent plasmon generators},}\ }\href@noop {} {\bibfield
   {journal} {\bibinfo  {journal} {Phys. Rev. B}\ }\textbf {\bibinfo {volume}
  {85}},\ \bibinfo {pages} {035448} (\bibinfo {year} {2012})}\BibitemShut
  {NoStop}%
\bibitem [{\citenamefont {Saito}\ \emph {et~al.}(1998)\citenamefont {Saito},
  \citenamefont {Dresselhaus},\ and\ \citenamefont
  {Dresselhaus}}]{saito1998physical}%
  \BibitemOpen
  \bibfield  {author} {\bibinfo {author} {\bibfnamefont {R.}~\bibnamefont
  {Saito}}, \bibinfo {author} {\bibfnamefont {G.}~\bibnamefont {Dresselhaus}},
  \ and\ \bibinfo {author} {\bibfnamefont {M.~S.}\ \bibnamefont
  {Dresselhaus}},\ }\href@noop {} {\emph {\bibinfo {title} {Physical properties
  of carbon nanotubes}}}\ (\bibinfo  {publisher} {Imperial College Press},\
  \bibinfo {address} {London},\ \bibinfo {year} {1998})\BibitemShut {NoStop}%
\bibitem [{\citenamefont {Helm}(2000)}]{weber1999intersubband}%
  \BibitemOpen
  \bibfield  {author} {\bibinfo {author} {\bibfnamefont {M.}~\bibnamefont
  {Helm}},\ }\bibfield  {title} {\enquote {\bibinfo {title} {The basic physics
  of intersubband transitions},}\ }in\ \href@noop {} {\emph {\bibinfo
  {booktitle} {Intersubband Transitions in Quantum Wells: Physics and Device
  Applications I}}},\ \bibinfo {series} {Semiconductors and Semimetals},
  Vol.~\bibinfo {volume} {62},\ \bibinfo {editor} {edited by\ \bibinfo {editor}
  {\bibfnamefont {H.~C.}\ \bibnamefont {Liu}}\ and\ \bibinfo {editor}
  {\bibfnamefont {F.}~\bibnamefont {Cappaso}}}\ (\bibinfo  {publisher}
  {Academic Press},\ \bibinfo {year} {2000})\ pp.\ \bibinfo {pages}
  {1--99}\BibitemShut {NoStop}%
\bibitem [{\citenamefont {Ajiki}\ and\ \citenamefont {Ando}(1993)}]{Ajiki1993}%
  \BibitemOpen
  \bibfield  {author} {\bibinfo {author} {\bibfnamefont {H.}~\bibnamefont
  {Ajiki}}\ and\ \bibinfo {author} {\bibfnamefont {T.}~\bibnamefont {Ando}},\
  }\bibfield  {title} {\enquote {\bibinfo {title} {Electronic states of carbon
  nanotubes},}\ }\href {\doibase 10.1143/jpsj.62.1255} {\bibfield  {journal}
  {\bibinfo  {journal} {J. Phys. Soc. Jpn.}\ }\textbf {\bibinfo {volume}
  {62}},\ \bibinfo {pages} {1255} (\bibinfo {year} {1993})}\BibitemShut
  {NoStop}%
\bibitem [{\citenamefont {Nakanishi}\ and\ \citenamefont
  {Ando}(2009)}]{Nakanishi2009}%
  \BibitemOpen
  \bibfield  {author} {\bibinfo {author} {\bibfnamefont {T.}~\bibnamefont
  {Nakanishi}}\ and\ \bibinfo {author} {\bibfnamefont {T.}~\bibnamefont
  {Ando}},\ }\bibfield  {title} {\enquote {\bibinfo {title} {Optical response
  of finite-length carbon nanotubes},}\ }\href {\doibase
  10.1143/jpsj.78.114708} {\bibfield  {journal} {\bibinfo  {journal} {J. Phys.
  Soc. Jpn.}\ }\textbf {\bibinfo {volume} {78}},\ \bibinfo {pages} {114708}
  (\bibinfo {year} {2009})}\BibitemShut {NoStop}%
\bibitem [{\citenamefont {Ritchie}(1957)}]{Ritchie1957}%
  \BibitemOpen
  \bibfield  {author} {\bibinfo {author} {\bibfnamefont {R.~H.}\ \bibnamefont
  {Ritchie}},\ }\bibfield  {title} {\enquote {\bibinfo {title} {Plasma losses
  by fast electrons in thin films},}\ }\href {\doibase 10.1103/physrev.106.874}
  {\bibfield  {journal} {\bibinfo  {journal} {Phys. Rev.}\ }\textbf {\bibinfo
  {volume} {106}},\ \bibinfo {pages} {874} (\bibinfo {year}
  {1957})}\BibitemShut {NoStop}%
\bibitem [{\citenamefont {Raether}(1980)}]{Raether1980}%
  \BibitemOpen
  \bibfield  {author} {\bibinfo {author} {\bibfnamefont {H.}~\bibnamefont
  {Raether}},\ }\href {\doibase 10.1007/bfb0045951} {\emph {\bibinfo {title}
  {Excitation of Plasmons and Interband Transitions by Electrons}}}\ (\bibinfo
  {publisher} {Springer-Verlag},\ \bibinfo {address} {Berlin Heidelberg},\
  \bibinfo {year} {1980})\BibitemShut {NoStop}%
\bibitem [{\citenamefont {Igarashi}\ \emph {et~al.}(2015)\citenamefont
  {Igarashi}, \citenamefont {Kawai}, \citenamefont {Yanagi}, \citenamefont
  {Cuong}, \citenamefont {Okada},\ and\ \citenamefont
  {Pichler}}]{Igarashi2015}%
  \BibitemOpen
  \bibfield  {author} {\bibinfo {author} {\bibfnamefont {T.}~\bibnamefont
  {Igarashi}}, \bibinfo {author} {\bibfnamefont {H.}~\bibnamefont {Kawai}},
  \bibinfo {author} {\bibfnamefont {K.}~\bibnamefont {Yanagi}}, \bibinfo
  {author} {\bibfnamefont {N.~T.}\ \bibnamefont {Cuong}}, \bibinfo {author}
  {\bibfnamefont {S.}~\bibnamefont {Okada}}, \ and\ \bibinfo {author}
  {\bibfnamefont {T.}~\bibnamefont {Pichler}},\ }\bibfield  {title} {\enquote
  {\bibinfo {title} {Tuning localized transverse surface plasmon resonance in
  electricity-selected single-wall carbon nanotubes by electrochemical
  doping},}\ }\href@noop {} {\bibfield  {journal} {\bibinfo  {journal} {Phys.
  Rev. Lett.}\ }\textbf {\bibinfo {volume} {114}},\ \bibinfo {pages} {176807}
  (\bibinfo {year} {2015})}\BibitemShut {NoStop}%
\bibitem [{\citenamefont {Ehrenreich}\ and\ \citenamefont
  {Cohen}(1959)}]{Ehrenreich1959}%
  \BibitemOpen
  \bibfield  {author} {\bibinfo {author} {\bibfnamefont {H.}~\bibnamefont
  {Ehrenreich}}\ and\ \bibinfo {author} {\bibfnamefont {M.~H.}\ \bibnamefont
  {Cohen}},\ }\bibfield  {title} {\enquote {\bibinfo {title} {Self-consistent
  field approach to the many-electron problem},}\ }\href@noop {} {\bibfield
  {journal} {\bibinfo  {journal} {Phys. Rev.}\ }\textbf {\bibinfo {volume}
  {115}},\ \bibinfo {pages} {786} (\bibinfo {year} {1959})}\BibitemShut
  {NoStop}%
\bibitem [{\citenamefont {Saito}\ \emph {et~al.}(2004)\citenamefont {Saito},
  \citenamefont {Gr{\"u}neis}, \citenamefont {Samsonidze}, \citenamefont
  {Dresselhaus}, \citenamefont {Dresselhaus}, \citenamefont {Jorio},
  \citenamefont {Can{\c{c}}ado}, \citenamefont {Pimenta},\ and\ \citenamefont
  {Filho}}]{Saito2004}%
  \BibitemOpen
  \bibfield  {author} {\bibinfo {author} {\bibfnamefont {R.}~\bibnamefont
  {Saito}}, \bibinfo {author} {\bibfnamefont {A.}~\bibnamefont {Gr{\"u}neis}},
  \bibinfo {author} {\bibfnamefont {Ge.~G.}\ \bibnamefont {Samsonidze}},
  \bibinfo {author} {\bibfnamefont {G.}~\bibnamefont {Dresselhaus}}, \bibinfo
  {author} {\bibfnamefont {M.~S.}\ \bibnamefont {Dresselhaus}}, \bibinfo
  {author} {\bibfnamefont {A.}~\bibnamefont {Jorio}}, \bibinfo {author}
  {\bibfnamefont {L.~G.}\ \bibnamefont {Can{\c{c}}ado}}, \bibinfo {author}
  {\bibfnamefont {M.~A.}\ \bibnamefont {Pimenta}}, \ and\ \bibinfo {author}
  {\bibfnamefont {A.~G.~Souza}\ \bibnamefont {Filho}},\ }\bibfield  {title}
  {\enquote {\bibinfo {title} {Optical absorption of graphite and single-wall
  carbon nanotubes},}\ }\href@noop {} {\bibfield  {journal} {\bibinfo
  {journal} {Appl. Phys. Lett.}\ }\textbf {\bibinfo {volume} {78}},\ \bibinfo
  {pages} {1099} (\bibinfo {year} {2004})}\BibitemShut {NoStop}%
\bibitem [{\citenamefont {Hertel}\ and\ \citenamefont
  {Moos}(2000)}]{Hertel2000}%
  \BibitemOpen
  \bibfield  {author} {\bibinfo {author} {\bibfnamefont {T.}~\bibnamefont
  {Hertel}}\ and\ \bibinfo {author} {\bibfnamefont {G.}~\bibnamefont {Moos}},\
  }\bibfield  {title} {\enquote {\bibinfo {title} {Influence of excited
  electron lifetimes on the electronic structure of carbon nanotubes},}\
  }\href@noop {} {\bibfield  {journal} {\bibinfo  {journal} {Chem. Phys.
  Lett.}\ }\textbf {\bibinfo {volume} {320}},\ \bibinfo {pages} {359} (\bibinfo
  {year} {2000})}\BibitemShut {NoStop}%
\bibitem [{\citenamefont {Reich}\ \emph {et~al.}(2002)\citenamefont {Reich},
  \citenamefont {Maultzsch}, \citenamefont {Thomsen},\ and\ \citenamefont
  {Ordej{\'{o}}n}}]{Reich2002}%
  \BibitemOpen
  \bibfield  {author} {\bibinfo {author} {\bibfnamefont {S.}~\bibnamefont
  {Reich}}, \bibinfo {author} {\bibfnamefont {J.}~\bibnamefont {Maultzsch}},
  \bibinfo {author} {\bibfnamefont {C.}~\bibnamefont {Thomsen}}, \ and\
  \bibinfo {author} {\bibfnamefont {P.}~\bibnamefont {Ordej{\'{o}}n}},\
  }\bibfield  {title} {\enquote {\bibinfo {title} {Tight-binding description of
  graphene},}\ }\href@noop {} {\bibfield  {journal} {\bibinfo  {journal} {Phys.
  Rev. B}\ }\textbf {\bibinfo {volume} {66}},\ \bibinfo {pages} {035412}
  (\bibinfo {year} {2002})}\BibitemShut {NoStop}%
\bibitem [{\citenamefont {Chegel}(2015)}]{Chegel2015}%
  \BibitemOpen
  \bibfield  {author} {\bibinfo {author} {\bibfnamefont {R.}~\bibnamefont
  {Chegel}},\ }\bibfield  {title} {\enquote {\bibinfo {title}
  {Third-nearest-neighbors tight-binding description of optical response of
  carbon nanotubes: Effects of chirality and diameter},}\ }\href@noop {}
  {\bibfield  {journal} {\bibinfo  {journal} {J. Electron. Mater.}\ }\textbf
  {\bibinfo {volume} {44}},\ \bibinfo {pages} {3500} (\bibinfo {year}
  {2015})}\BibitemShut {NoStop}%
\bibitem [{\citenamefont {Popov}(2004)}]{Popov2004}%
  \BibitemOpen
  \bibfield  {author} {\bibinfo {author} {\bibfnamefont {V.~N.}\ \bibnamefont
  {Popov}},\ }\bibfield  {title} {\enquote {\bibinfo {title} {Curvature effects
  on the structural, electronic and optical properties of isolated
  single-walled carbon nanotubes within a symmetry-adapted non-orthogonal
  tight-binding model},}\ }\href@noop {} {\bibfield  {journal} {\bibinfo
  {journal} {New J. Phys.}\ }\textbf {\bibinfo {volume} {6}},\ \bibinfo {pages}
  {17} (\bibinfo {year} {2004})}\BibitemShut {NoStop}%
\bibitem [{\citenamefont {Samsonidze}\ \emph {et~al.}(2003)\citenamefont
  {Samsonidze}, \citenamefont {Saito}, \citenamefont {Jorio}, \citenamefont
  {Pimenta}, \citenamefont {Souza~Filho}, \citenamefont {Gr{\"u}neis},
  \citenamefont {Dresselhaus},\ and\ \citenamefont
  {Dresselhaus}}]{samsonidze2003concept}%
  \BibitemOpen
  \bibfield  {author} {\bibinfo {author} {\bibfnamefont {Ge.~G.}\ \bibnamefont
  {Samsonidze}}, \bibinfo {author} {\bibfnamefont {R.}~\bibnamefont {Saito}},
  \bibinfo {author} {\bibfnamefont {A}~\bibnamefont {Jorio}}, \bibinfo {author}
  {\bibfnamefont {M.~A.}\ \bibnamefont {Pimenta}}, \bibinfo {author}
  {\bibfnamefont {A.~G.}\ \bibnamefont {Souza~Filho}}, \bibinfo {author}
  {\bibfnamefont {A.}~\bibnamefont {Gr{\"u}neis}}, \bibinfo {author}
  {\bibfnamefont {G.}~\bibnamefont {Dresselhaus}}, \ and\ \bibinfo {author}
  {\bibfnamefont {M.~S.}\ \bibnamefont {Dresselhaus}},\ }\bibfield  {title}
  {\enquote {\bibinfo {title} {The concept of cutting lines in carbon nanotube
  science},}\ }\href@noop {} {\bibfield  {journal} {\bibinfo  {journal} {J.
  Nanosci. Nanotechnol.}\ }\textbf {\bibinfo {volume} {3}},\ \bibinfo {pages}
  {431} (\bibinfo {year} {2003})}\BibitemShut {NoStop}%
\bibitem [{\citenamefont {Saito}\ \emph {et~al.}(2005)\citenamefont {Saito},
  \citenamefont {Sato}, \citenamefont {Oyama}, \citenamefont {Jiang},
  \citenamefont {Samsonidze}, \citenamefont {Dresselhaus},\ and\ \citenamefont
  {Dresselhaus}}]{Saito2005}%
  \BibitemOpen
  \bibfield  {author} {\bibinfo {author} {\bibfnamefont {R.}~\bibnamefont
  {Saito}}, \bibinfo {author} {\bibfnamefont {K.}~\bibnamefont {Sato}},
  \bibinfo {author} {\bibfnamefont {Y.}~\bibnamefont {Oyama}}, \bibinfo
  {author} {\bibfnamefont {J.}~\bibnamefont {Jiang}}, \bibinfo {author}
  {\bibfnamefont {Ge.~G.}\ \bibnamefont {Samsonidze}}, \bibinfo {author}
  {\bibfnamefont {G.}~\bibnamefont {Dresselhaus}}, \ and\ \bibinfo {author}
  {\bibfnamefont {M.~S.}\ \bibnamefont {Dresselhaus}},\ }\bibfield  {title}
  {\enquote {\bibinfo {title} {Cutting lines near the fermi energy of
  single-wall carbon nanotubes},}\ }\href@noop {} {\bibfield  {journal}
  {\bibinfo  {journal} {Phys. Rev. B}\ }\textbf {\bibinfo {volume} {72}},\
  \bibinfo {pages} {153413} (\bibinfo {year} {2005})}\BibitemShut {NoStop}%
\bibitem [{out()}]{outputPlasmon}%
  \BibitemOpen
  \href {link here} {}\bibinfo {note} {The plasmon data sets are open at
  \url{http://github.com/DariaSatco/PlasmonOutput}, while the absorption code
  is available at https://github.com/DariaSatco/cntabsorpt.}\BibitemShut {Stop}%
\bibitem [{\citenamefont {Sato}\ \emph {et~al.}(2007)\citenamefont {Sato},
  \citenamefont {Saito}, \citenamefont {Jiang}, \citenamefont {Dresselhaus},\
  and\ \citenamefont {Dresselhaus}}]{Sato2007}%
  \BibitemOpen
  \bibfield  {author} {\bibinfo {author} {\bibfnamefont {K.}~\bibnamefont
  {Sato}}, \bibinfo {author} {\bibfnamefont {R.}~\bibnamefont {Saito}},
  \bibinfo {author} {\bibfnamefont {J.}~\bibnamefont {Jiang}}, \bibinfo
  {author} {\bibfnamefont {G.}~\bibnamefont {Dresselhaus}}, \ and\ \bibinfo
  {author} {\bibfnamefont {M.~S.}\ \bibnamefont {Dresselhaus}},\ }\bibfield
  {title} {\enquote {\bibinfo {title} {Discontinuity in the family pattern of
  single-wall carbon nanotubes},}\ }\href {\doibase 10.1103/physrevb.76.195446}
  {\bibfield  {journal} {\bibinfo  {journal} {Phys. Rev. B}\ }\textbf {\bibinfo
  {volume} {76}},\ \bibinfo {pages} {195446} (\bibinfo {year}
  {2007})}\BibitemShut {NoStop}%
\bibitem [{\citenamefont {Nugraha}\ \emph {et~al.}(2010)\citenamefont
  {Nugraha}, \citenamefont {Saito}, \citenamefont {Sato}, \citenamefont
  {Araujo}, \citenamefont {Jorio},\ and\ \citenamefont
  {Dresselhaus}}]{Nugraha2010}%
  \BibitemOpen
  \bibfield  {author} {\bibinfo {author} {\bibfnamefont {A.~R.~T.}\
  \bibnamefont {Nugraha}}, \bibinfo {author} {\bibfnamefont {R.}~\bibnamefont
  {Saito}}, \bibinfo {author} {\bibfnamefont {K.}~\bibnamefont {Sato}},
  \bibinfo {author} {\bibfnamefont {P.~T.}\ \bibnamefont {Araujo}}, \bibinfo
  {author} {\bibfnamefont {A.}~\bibnamefont {Jorio}}, \ and\ \bibinfo {author}
  {\bibfnamefont {M.~S.}\ \bibnamefont {Dresselhaus}},\ }\bibfield  {title}
  {\enquote {\bibinfo {title} {Dielectric constant model for environmental
  effects on the exciton energies of single wall carbon nanotubes},}\ }\href
  {\doibase 10.1063/1.3485293} {\bibfield  {journal} {\bibinfo  {journal}
  {Appl. Phys. Lett.}\ }\textbf {\bibinfo {volume} {97}},\ \bibinfo {pages}
  {091905} (\bibinfo {year} {2010})}\BibitemShut {NoStop}%
\bibitem [{\citenamefont {Sato}\ \emph {et~al.}(2010)\citenamefont {Sato},
  \citenamefont {Saito}, \citenamefont {Nugraha},\ and\ \citenamefont
  {Maruyama}}]{Sato2010}%
  \BibitemOpen
  \bibfield  {author} {\bibinfo {author} {\bibfnamefont {K.}~\bibnamefont
  {Sato}}, \bibinfo {author} {\bibfnamefont {R.}~\bibnamefont {Saito}},
  \bibinfo {author} {\bibfnamefont {A.R.T.}\ \bibnamefont {Nugraha}}, \ and\
  \bibinfo {author} {\bibfnamefont {S.}~\bibnamefont {Maruyama}},\ }\bibfield
  {title} {\enquote {\bibinfo {title} {Excitonic effects on radial breathing
  mode intensity of single wall carbon nanotubes},}\ }\href {\doibase
  10.1016/j.cplett.2010.07.099} {\bibfield  {journal} {\bibinfo  {journal}
  {Chem. Phys. Lett.}\ }\textbf {\bibinfo {volume} {497}},\ \bibinfo {pages}
  {94} (\bibinfo {year} {2010})}\BibitemShut {NoStop}%
\bibitem [{\citenamefont {Samsonidze}\ \emph {et~al.}(2004)\citenamefont
  {Samsonidze}, \citenamefont {Saito}, \citenamefont {Kobayashi}, \citenamefont
  {Gr{\"u}neis}, \citenamefont {Jiang}, \citenamefont {Jorio}, \citenamefont
  {Chou}, \citenamefont {Dresselhaus},\ and\ \citenamefont
  {Dresselhaus}}]{Samsonidze2004}%
  \BibitemOpen
  \bibfield  {author} {\bibinfo {author} {\bibfnamefont {Ge.~G.}\ \bibnamefont
  {Samsonidze}}, \bibinfo {author} {\bibfnamefont {R.}~\bibnamefont {Saito}},
  \bibinfo {author} {\bibfnamefont {N.}~\bibnamefont {Kobayashi}}, \bibinfo
  {author} {\bibfnamefont {A.}~\bibnamefont {Gr{\"u}neis}}, \bibinfo {author}
  {\bibfnamefont {J.}~\bibnamefont {Jiang}}, \bibinfo {author} {\bibfnamefont
  {A.}~\bibnamefont {Jorio}}, \bibinfo {author} {\bibfnamefont {S.~G.}\
  \bibnamefont {Chou}}, \bibinfo {author} {\bibfnamefont {G.}~\bibnamefont
  {Dresselhaus}}, \ and\ \bibinfo {author} {\bibfnamefont {M.~S.}\ \bibnamefont
  {Dresselhaus}},\ }\bibfield  {title} {\enquote {\bibinfo {title} {Family
  behavior of the optical transition energies in single-wall carbon nanotubes
  of smaller diameters},}\ }\href {\doibase 10.1063/1.1829160} {\bibfield
  {journal} {\bibinfo  {journal} {Appl. Phys. Lett.}\ }\textbf {\bibinfo
  {volume} {85}},\ \bibinfo {pages} {5703} (\bibinfo {year}
  {2004})}\BibitemShut {NoStop}%
\bibitem [{\citenamefont {Hwang}\ and\ \citenamefont
  {Sarma}(2007)}]{hwang-graphene}%
  \BibitemOpen
  \bibfield  {author} {\bibinfo {author} {\bibfnamefont {E.~H.}\ \bibnamefont
  {Hwang}}\ and\ \bibinfo {author} {\bibfnamefont {S.~D.}\ \bibnamefont
  {Sarma}},\ }\bibfield  {title} {\enquote {\bibinfo {title} {Dielectric
  function, screening, and plasmons in two-dimensional graphene},}\ }\href@noop
  {} {\bibfield  {journal} {\bibinfo  {journal} {Phys. Rev. B}\ }\textbf
  {\bibinfo {volume} {75}},\ \bibinfo {pages} {205418} (\bibinfo {year}
  {2007})}\BibitemShut {NoStop}%
\bibitem [{\citenamefont {Sato}\ \emph {et~al.}(2017)\citenamefont {Sato},
  \citenamefont {Tatsumi},\ and\ \citenamefont {Saito}}]{Sato2017}%
  \BibitemOpen
  \bibfield  {author} {\bibinfo {author} {\bibfnamefont {N.}~\bibnamefont
  {Sato}}, \bibinfo {author} {\bibfnamefont {Y.}~\bibnamefont {Tatsumi}}, \
  and\ \bibinfo {author} {\bibfnamefont {R.}~\bibnamefont {Saito}},\ }\bibfield
   {title} {\enquote {\bibinfo {title} {Circular dichroism of single-wall
  carbon nanotubes},}\ }\href@noop {} {\bibfield  {journal} {\bibinfo
  {journal} {Phys. Rev. B}\ }\textbf {\bibinfo {volume} {95}},\ \bibinfo
  {pages} {155436} (\bibinfo {year} {2017})}\BibitemShut {NoStop}%
\bibitem [{\citenamefont {Pisarra}\ \emph {et~al.}(2016)\citenamefont
  {Pisarra}, \citenamefont {Sindona}, \citenamefont {Gravina}, \citenamefont
  {Silkin},\ and\ \citenamefont {Pitarke}}]{Pisarra2016}%
  \BibitemOpen
  \bibfield  {author} {\bibinfo {author} {\bibfnamefont {M.}~\bibnamefont
  {Pisarra}}, \bibinfo {author} {\bibfnamefont {A.}~\bibnamefont {Sindona}},
  \bibinfo {author} {\bibfnamefont {M.}~\bibnamefont {Gravina}}, \bibinfo
  {author} {\bibfnamefont {V.~M.}\ \bibnamefont {Silkin}}, \ and\ \bibinfo
  {author} {\bibfnamefont {J.~M.}\ \bibnamefont {Pitarke}},\ }\bibfield
  {title} {\enquote {\bibinfo {title} {Dielectric screening and plasmon
  resonances in bilayer graphene},}\ }\href {\doibase
  10.1103/physrevb.93.035440} {\bibfield  {journal} {\bibinfo  {journal} {Phys.
  Rev. B}\ }\textbf {\bibinfo {volume} {93}},\ \bibinfo {pages} {035440}
  (\bibinfo {year} {2016})}\BibitemShut {NoStop}%
\bibitem [{\citenamefont {Gomez}\ \emph {et~al.}(2016)\citenamefont {Gomez},
  \citenamefont {Pisarra}, \citenamefont {Gravina}, \citenamefont {Pitarke},\
  and\ \citenamefont {Sindona}}]{Gomez2016}%
  \BibitemOpen
  \bibfield  {author} {\bibinfo {author} {\bibfnamefont {C.~Vacacela}\
  \bibnamefont {Gomez}}, \bibinfo {author} {\bibfnamefont {M.}~\bibnamefont
  {Pisarra}}, \bibinfo {author} {\bibfnamefont {M.}~\bibnamefont {Gravina}},
  \bibinfo {author} {\bibfnamefont {J.~M.}\ \bibnamefont {Pitarke}}, \ and\
  \bibinfo {author} {\bibfnamefont {A.}~\bibnamefont {Sindona}},\ }\bibfield
  {title} {\enquote {\bibinfo {title} {Plasmon modes of graphene nanoribbons
  with periodic planar arrangements},}\ }\href@noop {} {\bibfield  {journal}
  {\bibinfo  {journal} {Phys. Rev. Lett.}\ }\textbf {\bibinfo {volume} {117}},\
  \bibinfo {pages} {116801} (\bibinfo {year} {2016})}\BibitemShut {NoStop}%
\bibitem [{\citenamefont {Torbatian}\ and\ \citenamefont
  {Asgari}(2017)}]{Torbatian2017}%
  \BibitemOpen
  \bibfield  {author} {\bibinfo {author} {\bibfnamefont {Z.}~\bibnamefont
  {Torbatian}}\ and\ \bibinfo {author} {\bibfnamefont {R.}~\bibnamefont
  {Asgari}},\ }\bibfield  {title} {\enquote {\bibinfo {title} {Plasmon modes of
  bilayer molybdenum disulfide: a density functional study},}\ }\href {\doibase
  10.1088/1361-648x/aa86b9} {\bibfield  {journal} {\bibinfo  {journal} {J.
  Phys. Condens. Matter}\ }\textbf {\bibinfo {volume} {29}},\ \bibinfo {pages}
  {465701} (\bibinfo {year} {2017})}\BibitemShut {NoStop}%
\bibitem [{\citenamefont {Torbatian}\ and\ \citenamefont
  {Asgari}(2018)}]{Torbatian2018}%
  \BibitemOpen
  \bibfield  {author} {\bibinfo {author} {\bibfnamefont {Z.}~\bibnamefont
  {Torbatian}}\ and\ \bibinfo {author} {\bibfnamefont {R.}~\bibnamefont
  {Asgari}},\ }\bibfield  {title} {\enquote {\bibinfo {title} {Plasmonic
  physics of $\mathrm{2D}$ crystalline materials},}\ }\href {\doibase
  10.3390/app8020238} {\bibfield  {journal} {\bibinfo  {journal} {Appl. Sci.}\
  }\textbf {\bibinfo {volume} {8}},\ \bibinfo {pages} {238} (\bibinfo {year}
  {2018})}\BibitemShut {NoStop}%
\end{thebibliography}
%

\end{document}